\documentclass[manuscript]{aastex61}

\usepackage{amsmath}
\usepackage{ulem}
\usepackage{color}
\usepackage{seqsplit}
\usepackage{hyperref}
\usepackage[figuresright]{rotating}
\usepackage{graphicx}

\newcommand{\simA}{D1-R7-IMF93}
\newcommand{\simB}{D2-R7-IMF01}

\newcommand{\msun}{M_\odot}

\begin{document}
\title{The long-term evolution of main-sequence binaries in DRAGON simulations}
\author{Qi Shu}
\affiliation{Department of Astronomy, School of Physics, Peking University, Yiheyuan Lu 5, Haidian Qu, 100871, Beijing, China}
\affiliation{Kavli Institute for Astronomy and Astrophysics, Peking University, Yiheyuan Lu 5, Haidian Qu, 100871, Beijing, China}

\author{Xiaoying Pang\footnote{Xiaoying.Pang@xjtlu.edu.cn}}
\affiliation{Department of Physics, Xi'an Jiaotong-Liverpool University, 111 Ren'ai Road, Suzhou Dushu Lake Science and Education Innovation District, Suzhou Industrial Park, Suzhou 215123, P.R. China}
\affiliation{Shanghai Key Laboratory for Astrophysics, Shanghai Normal University, 100 Guilin Road, Shanghai 200234, P.R. China}

\author{Francesco Flammini Dotti}
\affiliation{Department of Physics, Xi'an Jiaotong-Liverpool University, 111 Ren'ai Road, Suzhou Dushu Lake Science and Education Innovation District, Suzhou Industrial Park, Suzhou 215123, P.R. China}
\affiliation{Department of Physics, University of Liverpool, Liverpool L69 3BX, UK}

\author{M.B.N. Kouwenhoven}
\affiliation{Department of Physics, Xi'an Jiaotong-Liverpool University, 111 Ren'ai Road, Suzhou Dushu Lake Science and Education Innovation District, Suzhou Industrial Park, Suzhou 215123, P.R. China}

\author{Manuel Arca Sedda}
\affiliation{Astronomisches Rechen-Institut, Zentrum f\"ur Astronomie, University of Heidelberg, M\"onchhofstrasse 12--14, 69120, Heidelberg, Germany}

\author{Rainer Spurzem}
\affiliation{National Astronomical Observatories and Key Laboratory of Computational Astrophysics, Chinese Academy of Sciences, 20A Datun Rd., Chaoyang District, 100012, Beijing, China}
\affiliation{Astronomisches Rechen-Institut, Zentrum f\"ur Astronomie, University of Heidelberg, M\"onchhofstrasse 12--14, 69120, Heidelberg, Germany}
\affiliation{Kavli Institute for Astronomy and Astrophysics, Peking University, Yiheyuan Lu 5, Haidian Qu, 100871, Beijing, China}

\correspondingauthor{Xiaoying Pang}
\email{Xiaoying.Pang@xjtlu.edu.cn}

\submitjournal{ApJS and accepted on date 29 November 2020}
%Accepted for publication in ApJS on 29 November 2020
%%%%%%%%%%%%%%%%%%%%%%%%% 0-Abstract %%%%%%%%%%%%%%%%%%%%%%%%%
\begin{abstract}
We present a comprehensive investigation of main-sequence (MS) binaries in the DRAGON simulations, which are the first one-million particles direct $N$-body simulations of globular clusters. We analyse the orbital parameters of the binary samples in two of the DRAGON simulations, \simA{} and \simB{}, focusing on their secular evolution and correlations up to 12\,Gyr. These two models have different initial stellar mass functions: Kroupa 1993 (\simA{}) and Kroupa 2001 (\simB{}); and different initial mass ratio distributions: random paring (\simA{}) and a power-law (\simA{}).
In general, the mass ratio of a population of binaries increases over time due to stellar evolution, which is less significant in \simB{}. In \simA{}, primordial binaries with mass ratio $q \approx 0.2$ are most common, and the frequency linearly declines with increasing $q$ at all times. Dynamical binaries of both models have higher eccentricities and larger semi-major axes than primordial binaries. They are preferentially located in the inner part of the star cluster. Secular evolution of binary orbital parameters does not depend on the initial mass-ratio distribution, but is sensitive to the initial binary distribution of the system.
At $t=12$\,Gyr, the binary fraction decreases radially outwards, and mass segregation is present. A color difference of $0.1$~mag in $F330W-F814W$ and 0.2\,mag in $NUV-y$ between the core and the outskirts of both clusters is seen, which is a reflection of the binary radial distribution and the mass segregation in the cluster. The complete set of data for primordial and dynamical binary systems at all snapshot intervals is made publicly available.
\end{abstract}

%%%%%%%%%%%%%%%%%%%%%%%%% 1-Introduction %%%%%%%%%%%%%%%%%%%%
\section{Introduction} \label{sec:Introduction}

The vast majority of stars are thought to have formed as part of a binary or multiple stellar system \citep[e.g.,][and references therein]{Shatsky2002, kouwenhoven2005, duchene2013}. Most known binary systems have been discovered in radial velocity surveys, transit surveys, and direct imaging surveys. 
It is estimated that less than one-third of the nearby solar-type field stars is single \citep{2004A&A...415..391M, 2003ApJ...586..512B, 2010ApJS..190....1R}.
The multiplicity fraction among field stars increases along with the primary mass \citep[e.g.,][and references therein]{2007prpl.conf..133G, duchene2013},
ranging from $\sim 22\%$ for M-dwarfs \citep[e.g.,][]{allen2007} to nearly 100\% for OB stars \citep[e.g,][]{Shatsky2002, kobulnicky2006, sana2011, 2012Sci...337..444S}. 
Observational surveys have also found that open clusters hold a significant higher binary fraction of 35\% up to 70\% \citep[e.g.,][]{2010MNRAS.401..577S} than globular clusters, which typically have binary fractions of $3-38$\% \citep[e.g.,][]{2012A&A...540A..16M}.

Open clusters, globular clusters, OB associations, and star forming regions provide excellent laboratories for studying stellar evolution and binary evolution \citep[e.g.,][]{kalirai2010, catelan2010}. Similarly, binary systems provide indispensable tools for studying the historical and present-day properties of these stellar groupings. 
The vast majority of binaries in most star clusters, notably globular clusters, remain unresolved in direct imaging surveys, due to their large distances, which results in crowding, while individual stars are normally too faint for radial velocity surveys.  The most efficient method for constraining the global properties of the binary population in massive star clusters makes use of the colour-magnitude diagram that enables the identification of unresolved binaries that populate certain regions in the diagram \citep[][]{2010MNRAS.401..577S, 2012A&A...540A..16M, 2016MNRAS.455.3009M}.

Despite substantial observational and theoretical efforts, the process of star formation is still not fully understood. The primordial binary population is a direct outcome of star formation, and can therefore provide important constraints about the star formation process \citep[e.g.,][]{kroupa1995, marks2012}. The origin of these primordial binary stars is strongly related to conservation and dissipation of angular momentum as proto-stellar clouds contract \citep[see, e.g.,][]{larson2003, mckee2007}. Interaction of the stars in a binary with a circumstellar disk tends to lead to more or less equal-mass binary systems \citep[e.g.,][]{tokovinin2000, krumholz2007}, while the fragmentation of massive circumstellar disks tends to result in unequal-mass, lower-mass binary systems \cite[e.g.,][]{li2015, li2016}. The formation process of primordial binary systems is highly complex and may be strongly affected by the physical properties of the molecular cloud and by the presence of strong radiation fields, magnetic fields \citep[e.g.,][]{price2007}, and close encounters with nearby stars \citep[e.g.,][]{whitworth2001}.

The orbital elements of binary stars may gradually evolve over time, and some of the binaries are completely disrupted through either internal or external processes. This may occur as a consequence of dynamical interactions with neighbouring stars \citep[see, e.g.,][and references therein]{1975MNRAS.173..729H, hut1992}, or in the case of very wide binary systems, by the external galactic tidal field \citep[e.g.,][]{jiang2010}. Stellar evolution and mass loss in detached binary systems, as well as binary evolution, will affect binary stars over time \citep{Hurley:2002ab}. Secular evolution in multiple stellar system can have similar effects \citep[e.g.,][]{ford2000, naoz2013, hamers2020}.

Over time, new binary stars may form as a result of dynamical interactions. This process is very inefficient in the galactic field \citep[e.g.,][]{1993ApJ...403..271G}, but three-body interactions can lead to the  formation of population of dynamical binaries in star clusters \citep[e.g.,][]{2017MNRAS.468.2429B}. The direct formation of close binary systems in star cluster is rare, although hardening of binaries due to interaction with neighbour stars occasionally results in short-period dynamical binary systems. Star clusters tend to have a transient population of very wide binary systems \citep[][]{Moeckel:2011aa} that may freeze in as the star cluster evolves \citep[e.g.,][]{Kouwenhoven:2010aa} and contribute to the wide binaries in the Galactic field.

Accurate knowledge of the present-day properties of the binary population provides comprehensive information about the formation history and the dynamical evolution the stellar population. The amount of information provided by the binary population is enormous, which poses opportunities as well as major challenges. Firstly, it is difficult to quantify the available information in the multi-dimensional parameter space using simple expressions, such as a binary fraction or a prescription for the mass ratio distribution, without significant loss of relevant information. Secondly, observational limitations have thus far limited the astronomical community from providing a complete record of the stellar population, even in the neighbourhood of the Sun, although substantial progress has been made in recent years \citep[see, e.g.,][and references therein]{tokovinin2014a, tokovinin2014b}.

The  binary population in very young and sparse stellar groupings \citep[such as OB associations, e.g.,][]{2010ApJS..190....1R} are the least affected by stellar evolution and binary evolution, and is therefore closest to the outcome of the star formation process. OB associations and star forming regions are often used to relate the properties of the binary population to the star forming process. The binary population in globular clusters, on the other hand, is strongly affected by both stellar evolution and dynamical evolution. Globular clusters are bright, abundant, and can be studied in detail at intergalactic distances. The observable properties of globular clusters can provide important information on both stellar evolution and dynamical evolution, and at the same time, can provide constraints on the primordial binary population, at substantially earlier epochs, and at much lower metallicities.

Characterizing the binary population in globular clusters is challenging, as is constraining the primordial binary population. Unlike in nearby star-forming regions, it is substantially more challenging to survey individual stars for binary companions in  extra-galactic globular clusters, even though obtaining multi-band photometry and spectroscopic measurements is not difficult. In addition, it is possible to obtain measurements of distinguishing features of binaries. Even though it may not be possible to observe binaries directly, such measurements can be used to constrain the properties of the binary population (including the primordial binary population), and to exclude alternatives.

Direct $N$-body simulations can be model to study star clusters and their constituent populations. Modern $N$-body simulations are able to model the relevant physical processes that occur in star clusters, most importantly stellar dynamics,  stellar evolution, and the influence of external tidal fields \citep[see, e.g.,][]{Aarseth:1999aa}. Developments in both software and hardware have significantly improved in the recent decade, resulting in significant speed-ups, as well as the ability to model star clusters with unprecedented accuracy. The recent implementation of the use of Graphical Processing Units (GPUs) and hybrid parallelization methods made the direct $N$-body simulations for globular clusters feasible \citep{Wang:2015aa}.

Substantial progress was made with the DRAGON simulation project \citep{Wang:2016aa}. 
The DRAGON simulations are the first 1-million particle direct $N$-body simulations of globular clusters. The set of DRAGON simulations consists of four models. Each model is initialized with a binary fraction of 5\%, and were dynamically evolved using NBODY6++GPU \citep{Wang:2015aa, Wang:2016aa}. 
In these simulations, the evolution the stellar and binary population, can be modeled more accurately than those carried out using in Monte Carlo simulations \citep{Wang:2016aa}, albeit at a greater computational cost. Moreover, the properties of the binary population
are recorded frequently, at an interval of 0.1\,Myr, such that these can be easily compared with observations, to deepen our understanding of the physical processes involved. In these study we distinguish between primordial binaries and dynamical binaries. We define a primordial binary system as a a binary system as a binary system that were present at the start of the simulation. Dynamical binary systems, on the other hand, are binary systems that form at a later time, through capture or exchange.

There are four models in the DRAGON simulation project: \simA{}, \simB{}, D3-R7-ROT and D4-R3-IMF01.
We carry out a detailed investigation on the long-term evolution of the properties of stellar binary systems in the DRAGON simulations, and compare our findings with properties of observed globular clusters, in order to provide the underlined physics for observed features in current and future observations. We will focus primarily on the parameter characterization of main-sequence (MS) binaries, which are by far the most common types of binary systems in star clusters,  and in the Galactic field. The analysis of the temporal  evolution of the orbital elements of the binary systems provides information about the formation, destruction, and secular evolution of binary systems, whereas comparing these with their location inside the parent cluster will provide us with insights on dynamical processes like mass segregation.
Distributions of primordial and dynamical binaries 
are compared over time and position, in order to evaluate whether our result are in agreement with realistic globular cluster data. We will compare the secular evolution of orbital elements, and mass ratio, of the two different binary types, using the probability distribution (i.e., the relative frequency) for a large set of data (e.g., the primordial binaries), while we use the cumulative distribution for smaller sets of data (e.g., the dynamical binaries).
Subsequently, we will study the correlations between the orbital elements and the correlations with mass ratio, in order to find how binaries mass ratio groups distribute in their orbital parameters.
Finally, we carry out an observational comparison with our data, in order to analyse how they compare with the results of our simulations.
We also provide an online database with the kinematic and orbital properties of all binary systems in the two DRAGON models analysed in this study.

This paper is organized as follows. In Section~\ref{sec:binary_sample} we present the initial conditions of the binary systems in the DRAGON simulations. In Section~\ref{sec:binary_stats} we analyse the evolution of the binary population over time, and identify correlations between different orbital properties, physical properties, and dynamical histories. We provide an observational analysis in Section~\ref{sec:dynamics_signatures}. Finally, we summarize and discuss our findings in Section~\ref{sec:summary}.

%%%%%%%%%%%%%%%%%%%%%%%%% 2-binary_sample %%%%%%%%%%%%%%%%%%%
\section{Binary samples from DRAGON simulations}\label{sec:binary_sample}
\subsection{Initial conditions of DRAGON simulations}\label{sec:DRAGON_simulation}

Among four models of DRAGON simulations, model D4-R3-IMF01 was initialized with a more compact density profile. This model was evolved for only 1\,Gyr, which is too short for a comparison with observed globular clusters. Cluster model D3-R7-ROT included global rotation. However, it did not show any significantly different results in the evolution of the binary population, which is the focus of our study. The other two models, \simA{} and \simB{}, have many similar initial conditions and a few differences. The small differences between both models result in a different evolution that deserves a comparison. Therefore, in this work we select \simA{} and \simB{} as our target models to study the evolution of the binary population in detail.
Both models aim to simulate the dynamical evolution of globular cluster NGC\,4372. In both models, the globular cluster is placed in an external tidal field corresponding to that of the Milky Way, at NGC\,4372's Galactocentric distance of 7.1\,kpc. The star cluster is placed in a circular orbit around the Milky Way centre. The Galactic centre is represented with a point mass, corresponding to the enclosed mass of the Milky Way of $8 \times 10^{10}~\msun$ and exerting the tidal force on the star cluster. In both models, the clusters are initially tidally under-filled.

As direct $N$-body simulations, the DRAGON simulations represent each particle as an individual star in the cluster.  These globular cluster models are initialized with 950,000 single stars and 50,000 primordial binary systems.
We list the main properties of the two models in  Table~\ref{table:initail}.
Both globular cluster models have a \cite{1966AJ.....71...64K} initial density profile with a King dimensionless parameter $W_0=6$, and are assigned initial half-mass radii, $R_{h,0}$, of 7.5~pc for \simA{} and 7.6~pc \simB.
Both globular clusters are initialized in virial equilibrium, $Q=0.5$, where the virial ratio $Q=|T/U|$ is the ratio between the total kinetic energy ($T$) and the total potential energy ($U$) of the star cluster.

One major difference between the initial conditions of \simA{} and \simB{} is the initial mass function (IMF).
Model \simA{} adopts the IMF of \cite{1993MNRAS.262..545K}, while \simB{} uses that of \cite{2001MNRAS.322..231K}. The IMF is initialised following $f(m) \propto m ^{-\alpha}$ in the mass range $0.08-100\msun$.
Both simulations adopt $\alpha = 1.3$ for the mass range $0.08 < m/\msun \le 0.5$. 
The value of $\alpha$ differs in $0.5 < m/\msun \le 1$, with $\alpha = 2.2$ for \simA, and $\alpha = 2.3$ for model \simB. 
Finally, for $m > 1 \ \msun$ range, \simA{} uses $\alpha = 2.7$ and \simB{} uses $\alpha = 2.3$. 
Due to different slopes in the IMFs, \simB{} has a higher total mass than \simA. Therefore it also has a slightly larger tidal radius ($R_{t,0}$), as shown in Table~\ref{table:initail}.
The initial metallicity of both models, $Z=0.00016$ was chosen identical to that of the globular cluster NGC~4372 \citep{1995AJ....109..605G, 2014A&A...567A..69K, 2015A&A...579A...6S}.

Each model is initialized with a binary fraction of 5\%. The start of the simulations ($t=0$) represents the time after which the star clusters have expelled all their interstellar gas, have smothered out their initial substructure, and have obtained virial equilibrium. Moreover, soft binaries are assumed to have been destroyed; the binary population at $t=0$~Myr is thus modeled to represent the hard binary population at that epoch. In the analysis below, the first snapshot (0.1\,Myr) after the start of the simulations is defined as $t=0$~Myr. The initial semi-major axes of the primordial binary system follows a log-normal distribution in the range $0.005-50$~AU.
The lower limit is a consequence of the physical size of the stars, while the upper limit is near the hard-soft boundary for stars in these globular clusters. The initial eccentricity distribution is thermal: $f(e) \propto e$ \citep[see, e.g.,][]{1975MNRAS.173..729H, 1993ApJ...403..271G}.

The initial mass ratio distribution of the binary stars, $f(q)$, is also different for the two simulations. Here, the mass ratio is defined as $q=m_2/m_1$, where $m_1$ is the mass of the primary star, and $m_2$ is the mass of the companion star (i.e., the least massive of the two stars in the binary system). 
In \simA, the mass of primordial binaries were generated by random pairing from the IMF. 
In \simB, the mass of the primary star ($m_1$) is randomly chosen from the IMF, and the secondary mass $m_2=q m_1$ is subsequently assigned after drawing the mass ratio from the probability distribution $f(q) \propto q^{-0.4}$ \citep{Kouwenhoven2007aa} in the mass range $0.08 \msun \leq m_2 \leq 100 \msun$. The latter lower limit ensures that both primary stars and companions stars are in the mass range $0.08-100 \msun$.  
This major difference will lead to variation in binary evolution (see Section~\ref{sec:mass_ratio_evolution}).

Stellar evolution is modeled following the prescriptions of \cite{Hurley:2000aa} for single stars, and those of \cite{Hurley:2002ab} for interacting binary systems.
The gravitational kicks exerted on neutron stars and black holes follow a Maxwellian velocity distribution. The variance of the Maxwellian distribution in model \simA{} differs from that in model \simB{} (see Table 1).
The kick velocity determines the number of remaining bound neutron stars and the retained black hole subsystems in the cluster, which affects the global dynamics of the cluster.

\begin{table}
	\centering
	\caption{
	Initial conditions for the two DRAGON simulations studied in this paper.
	$Z$ is metallicity value of the cluster;
	$Q$ is initial virial ratio of the cluster;
	$B$ is initial binary fraction of the cluster;
	$f(q)$ denotes the initial binary mass ratio distribution; 
	$f(e)$ and $f(a)$ are initial eccentricity and semi-major axis distribution of primordial binaries; 
	$R_{h,0}$ is the initial half-mass radius; 
	$R_{t,0}$ is the initial tidal radius;
	$\sigma_k$ characterises the Maxwellian distribution for gravitational kick velocities.
	This table shows the similar initial condition information as Table 1 of \cite{Wang:2016aa}.}
	\label{table:initail}	
	\begin{tabular}{lll} 
		\hline
		         &  \simA{} & \simB{} \\
		\hline
		        Profile & \cite{1966AJ.....71...64K} $W_0$ = 6  & \cite{1966AJ.....71...64K} $W_0$ = 6  \\
		        M ($\msun$) & 474\ 603 & 591\ 647 \\
				IMF  & \cite{1993MNRAS.262..545K} & \cite{2001MNRAS.322..231K}  \\
				$Z$ & 0.00016  & 0.00016 \\ 
				$Q$ & 0.5 & 0.5 \\
				$B$ & 5\% & 5\% \\
				$f(q)$ & random pairing & \cite{Kouwenhoven2007aa} \\
				$f(e)$ & thermal, $\propto e$ & thermal, $\propto e$ \\
				$f(a)$ (AU) & logarithmic normal & logarithmic normal \\
				$R_{h,0}$  & 7.5 pc & 7.6 pc    \\
				$R_{t,0}$  & 89 pc & 97 pc                  \\
        		{\bf $\sigma_k$}  & 30 km\,s$^{-1}$ & 265 km\,s$^{-1}$   \\
		\hline
	\end{tabular}
\end{table}
%%%%%%%%%%%%%%%%%%%%%%%%%%%%%%%%%%%%%%%%%%%%%%%%

\subsection{Binary star data}

Primordial binaries are present immediately after the star formation process has ended \citep[e.g.,][]{2003IAUS..221P..49K}, and have therefore, in an idealised scenario, not yet been affected by dynamical evolution. Over time, existing binary systems are disrupted, and new binary systems are formed.
Dynamical binary systems are formed as the simulation proceeds, through dynamical close encounters between single stars and/or binary systems. The two stars that form the new binary system may or may not have been part of a primordial (and/or dynamical binary) at earlier times. In this paper, We consider the binaries present in the initial conditions of \simA{} and \simB{} as primordial binaries, which can be identified with the unique IDs. Any binaries formed at later times are considered as dynamical binaries, whose IDs are different from the primordial ones.

We identify all binary systems in each snapshot of both models, for a period of 12~Gyr, and identify the evolution of each binary system. The evolution of total number of MS stars, $N$, the properties of MS binaries for both simulations are listed in Tables~\ref{table:binary_number_R7_B5_solar} and~\ref{table:binary_number_R7_IMF2001_RG7.1}, for 
snapshots at times 0  (more precisely, 0.1), 100, 300, 600, 1000, 3000, 6000, and 12000~Myr. The detailed binary samples for each snapshot are provided in Appendix~\ref{section:appendixA}.

\begin{table*}
	\centering
	\caption{Numerical evolution of the number of MS binaries in \simA, at different times. Binary fraction is defined as the total number of binaries divided by the total number of binaries and single stars. Time, in units of Myr, is indicated in the first row. } 
	\label{table:binary_number_R7_B5_solar}	
	\begin{tabular}{lc rrrrrrrr} 
		\hline
		Property & & 0 & 100 & 300 & 600 &1000& 3000& 6000& 12000 \\
		\hline
    All particles      & $N$  &  1\,050\,000 &  1\,046\,536 &  1\,045\,610 &  1\,044\,067 &  1\,041\,297 &  1\,020\,529 &   986\,958 &   923\,860 \\
    Binary fraction (\%)   & $\mathcal{B}$ &   4.73 &   4.63 &   4.50 &   4.36 &   4.21 &   3.84 &   3.55 &   3.31 \\
    All binaries        & $B$  &    49\,628 &    48\,458 &    47\,036 &    45\,489 &    43\,885 &    39\,169 &    35\,007 &    30\,568 \\
    Primordial binaries & $B_P$  &    49\,628 &    48\,446 &    47\,018 &    45\,454 &    43\,833 &    39\,059 &    34\,860 &    30\,359 \\
    Dynamical binaries  & $B_D$  &        0 &       12 &       18 &       35 &       52 &      110 &      147 &      209 \\
		\hline
	\end{tabular}
\end{table*}

\begin{table*}
	\centering
	\caption{Numerical evolution of MS binaries in \simB, at different times. The parameters are the same as those in Table~\ref{table:binary_number_R7_B5_solar}. 
	}
	\label{table:binary_number_R7_IMF2001_RG7.1}	
	\begin{tabular}{lc rrrrrrrr} 
		\hline
		Property & & 0 & 100 & 300 & 600 &1000& 3000& 6000& 12000 \\
		\hline
    All particles      & $N$  &  1\,050\,000 &  1\,041\,914 &  1\,039\,032 &  1\,032\,165 &  1\,021\,608 &   966\,365 &   887\,621 &   729\,738 \\
    Binary fraction (\%)   & $\mathcal{B}$ &   4.76 &   4.72 &   4.61 &   4.48 &   4.39 &   4.13 &   3.94 &   3.77 \\
    All binaries        & $B$  &    50\,003 &    49\,151 &    47\,880 &    46\,252 &    44\,798 &    39\,894 &    34\,935 &    27\,530 \\
    Primordial binaries & $B_P$  &    49\,997 &    49\,136 &    47\,857 &    46\,223 &    44\,756 &    39\,831 &    34\,840 &    27\,438 \\
    Dynamical binaries  & $B_D$  &        6 &       15 &       23 &       29 &       42 &       63 &       95 &       92 \\

		\hline
	\end{tabular}
\end{table*}
%\clearpage

%%%%%%%%%%%%%%%%%%%%%%%%% 3-binary_stats %%%%%%%%%%%%%%%%%
\section{Secular evolution of binary properties}\label{sec:binary_stats}
In this section we investigate the secular evolution of binary parameters, such as the mass ratio, the primary mass, the semi-major axis, and the eccentricity. We also discuss the statistical significance of the properties of the primordial and dynamical binary systems in models \simA{} and \simB{}. This study focuses on MS binaries, i.e., systems in which both  the primary and secondary are MS stars. 

\subsection{Mass ratio}
\label{sec:mass_ratio_evolution}

%%%%%%%%%%%%%%%%%%%%%%%%%%%%%%%%%%%%%%%%%%%%%%%%%%%%%%%%
%%%%%%%%%%%%%%%
% mass ratio
%%%%%%%%%%%%%%%

\begin{figure*}
  \centering
  \includegraphics[width=0.9\textwidth]{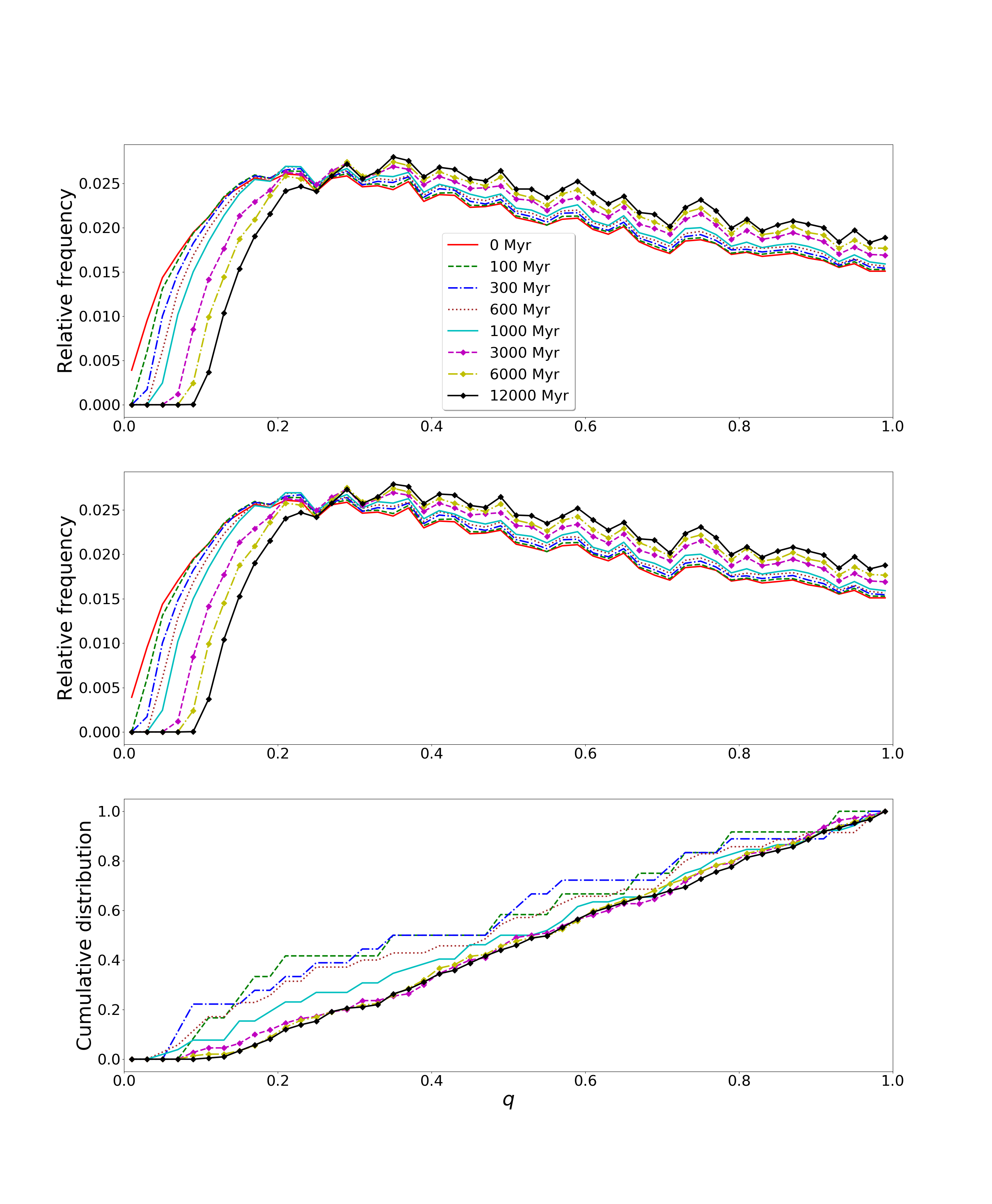}
  \caption{Evolution of the mass ratios of MS binaries in model \simA. The panels show normalized mass ratio distribution, $f(q)$, for all binaries ({\em top}), the normalized mass ratio distribution for the primordial binaries ({\em middle}), and the cumulative distribution of dynamical binaries ({\em bottom}). Different colors and symbols indicate different times. 
}\label{fig:mass_ratio_evolution_R7_B5_solar}
\end{figure*}

\begin{figure*}
  \centering
  \includegraphics[width=0.9\textwidth]{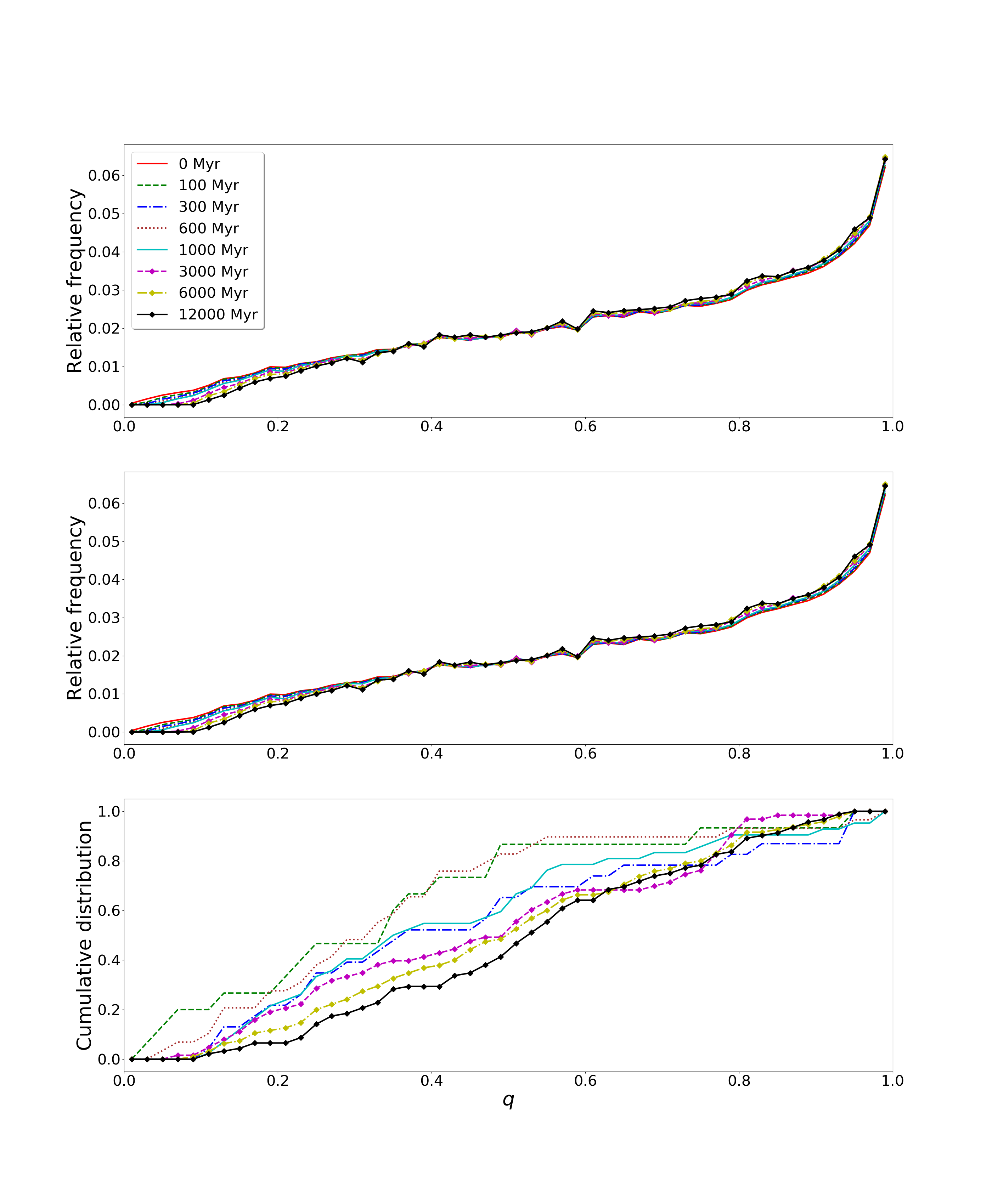}
  \caption{Same as in Figure~\ref{fig:mass_ratio_evolution_R7_B5_solar}, but for model \simB.
  }\label{fig:mass_ratio_evolution_R7_IMF2001_RG71}
\end{figure*}

 Figure~\ref{fig:mass_ratio_evolution_R7_B5_solar} shows the secular evolution of MS binary mass ratio in model \simA. Throughout this work, we define the mass ratio, $q=m_2/m_1$, as the ratio of the masses of the secondary star, $m_2$, and the primary star, $m_1$, where $m_1 \geq m_2$.
 Binary systems in model \simA{} are generated through random pairing of the two components, which are independently drawn from the IMF. The mass ratio distribution for the combined sample of primordial MS binaries generated through random pairing is highest in the range $0.2<q<0.4$, and remains so during the entire simulation. The relative frequency of binary systems with $q \le 0.2$ decreases over time. At early times, primary stars of binaries with $q \la 0.2$ are typically more massive than those of binaries with $q \ga 0.2$.
 The former evolve faster and experience higher mass loss, resulting in a tendency of the average mass ratio of the MS binary population to increase over time. 
 
The relative frequency of MS binaries $q \le 0.2$ in \simB{} also decreases over time (see Figure~\ref{fig:mass_ratio_evolution_R7_IMF2001_RG71}), similar to in model \simA{}. However, the decrease is less prominent due to the different initial mass ratio distribution of \cite{Kouwenhoven2007aa} that is assigned to the population. For MS binaries in model \simB, the initial fraction of MS stars with $q \le 0.2$ in the combined sample is lower than that for model \simA. Model \simA{} has a higher fraction of MS binaries with lower mass ratios, and therefore more massive MS binaries. 
 
 Throughout the entire simulation, the number of dynamically-formed MS binaries remains more than two orders of magnitude smaller than the number of primordial MS binaries, for both model \simA{} and model \simB{}. The bottom panels of Figures~\ref{fig:mass_ratio_evolution_R7_B5_solar} and~\ref{fig:mass_ratio_evolution_R7_IMF2001_RG71} show the mass ratio distributions for the dynamical binary systems in both models. Note that, while in the top and middle panels in these figures we show mass ratio distributions, $f(q)$, the bottom panels show cumulative mass ratio distributions, $F(q)$, which are more suitable for representing the much smaller number of dynamical binary systems. Note that the mass ratio distribution of a population of binary systems with a restricted primary star mass range may look substantially different from that of the entire population of binary systems \citep[see, e.g.,][and references therein]{Kouwenhoven2009aa}. 

 %%%%%%%%%%%%%%%
\begin{figure}
  \centering
  \includegraphics[width=\columnwidth]{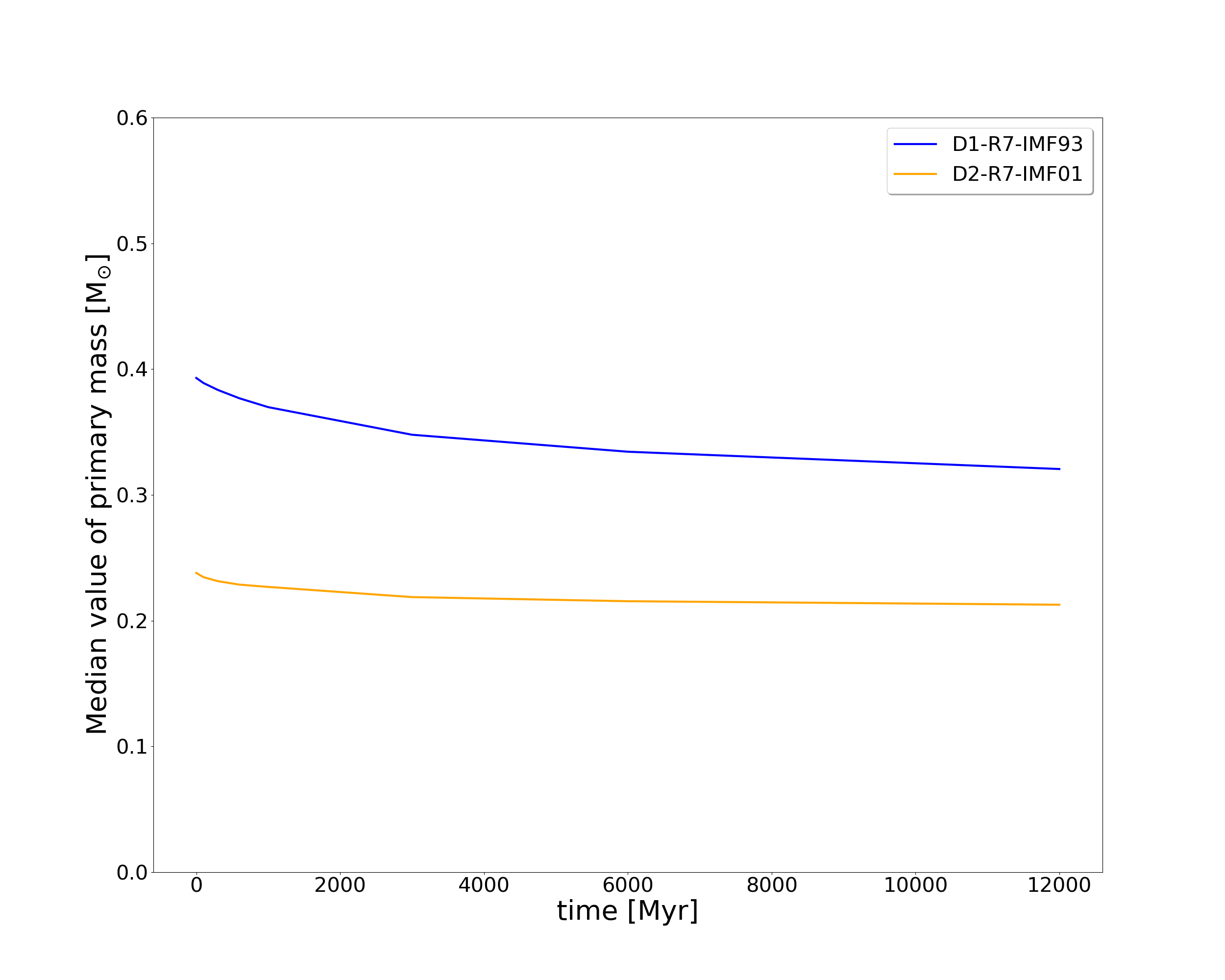}
  \caption{Secular evolution of the median value of primary mass for model \simA{} (blue curve) and model \simB{} (orange curve).}
  \label{fig:median_value_primary_mass_evolution}
\end{figure}

Figure~\ref{fig:median_value_primary_mass_evolution} shows the secular evolution of the median primary masses of the MS binaries in models \simA{} and \simB.  The median primary mass resulting from random paring (\simA) is generally larger than that resulting from the \cite{Kouwenhoven2007aa} recipe (model \simB), fully consistent with Figures~\ref{fig:mass_ratio_evolution_R7_B5_solar} and~\ref{fig:mass_ratio_evolution_R7_IMF2001_RG71}. Over time, the median primary mass in both models decreases as a consequence of stellar evolution.

 Numerous observational studies have made efforts to obtain parameter distributions of binary systems in star clusters \citep[e.g.,][]{2010MNRAS.401..577S, 2012A&A...540A..16M, 2016MNRAS.455.3009M}. In addition to the binary fraction, the most sought-after parameter parameter distributions are that of the mass ratio and that of the semi-major axis (or orbital period). The mass ratio distribution of massive MS binaries (B-type), in the OB-association Sco~OB2, is well-described by the power law $f(q)\propto q^{-0.5}$ \citep{Shatsky2002}, similar to \cite{Kouwenhoven2007aa}.  The mass ratio distribution in the regime $q\leq0.1$ of these massive binaries in Sco~OB2, is similar to our binary samples at the time before 1~Gyr.   On the other hand, a flat distribution of the mass ratio (for $q > 0.5$) has been observed in 59 Galactic globular clusters \citep[see][]{2012A&A...540A..16M}.

\cite{2010ApJS..190....1R} carried out a systematic study for binary companions amongst 454 solar-type stars the Galactic neighbourhood.  The mass ratio distribution of these field binaries peaks at $q\approx 1$  \citep{2010ApJS..190....1R}, with companion star masses range in the range $0.05-0.98\msun$. 
\simB{} appears to be consistent with \citet{2010ApJS..190....1R}. However, if the high-order binaries (i.e., triples or quadruples) are removed in \citet{2010ApJS..190....1R}, the mass ratio distribution flattens.

%%%%%%%%%%%%%%%%%%%%%%%%%%%%%%%%%%%%%%%%%%%%%
%%%%%%%%%%%%%%%
% semi-major axis
%%%%%%%%%%%%%%%

\subsection{Semi-major axis}
\label{sec:semi_major_axis_evolution}

\begin{figure*}
  \centering
  \includegraphics[width=0.97\textwidth]{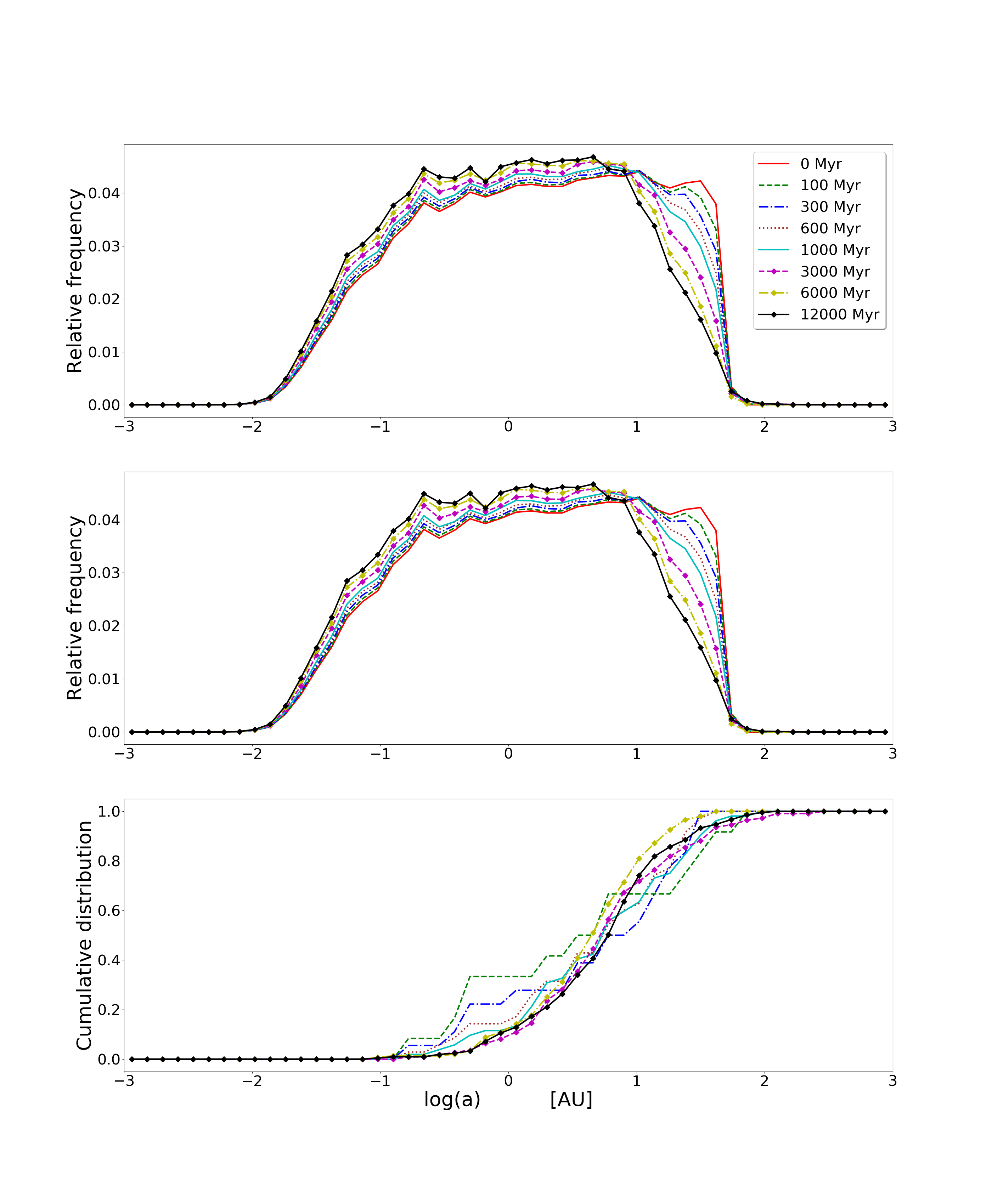}
  \caption{Secular evolution of the semi-major axis $a$ of MS binaries in \simA. The colors and symbols are identical to those in Figure~\ref{fig:mass_ratio_evolution_R7_B5_solar}. }\label{fig:semi_evolution_R7_B5_solar}
\end{figure*}

\begin{figure*}
  \centering
  \includegraphics[width=0.97\textwidth]{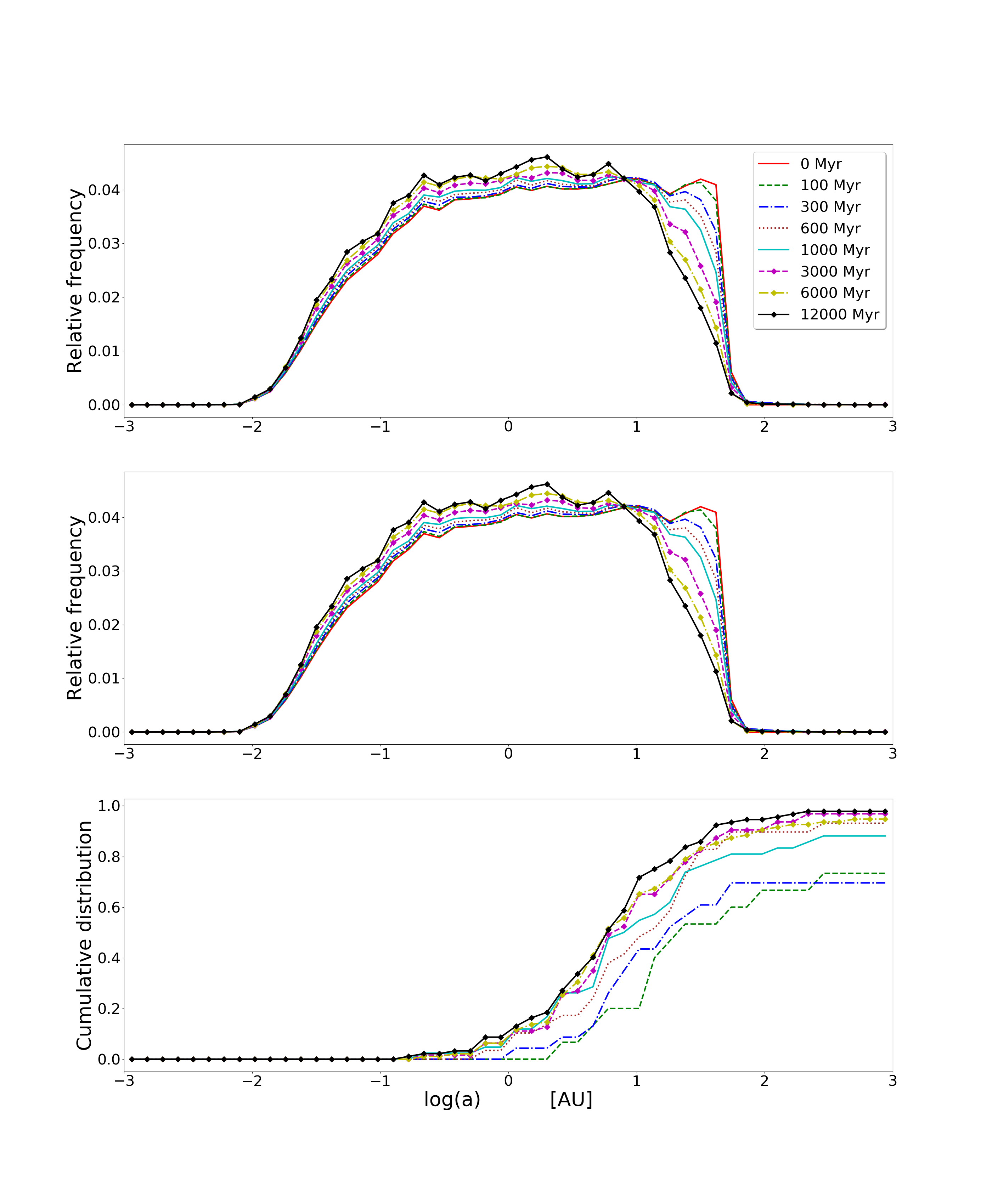}
  \caption{Secular evolution of the semi-major axis $a$ of MS binaries in \simB. The colors and symbols are identical to those in Figure~\ref{fig:mass_ratio_evolution_R7_B5_solar}.  }\label{fig:semi_evolution_R7_IMF2001_RG71}
\end{figure*}

The semi-major axis, $a$, is another important parameter in binary evolution. We show the secular evolution of the semi-major axis distribution, $f(a)$, in 
Figure~\ref{fig:semi_evolution_R7_B5_solar} (for \simA) and Figure~\ref{fig:semi_evolution_R7_IMF2001_RG71} (for \simB).
The initial semi-major axis distribution is log-normal, ranging from $a=0.005$~AU to $a=50$~AU for both simulations. Binary systems with large semi-major axis are easily disrupted, since they are vulnerable to close encounters with neighbouring stars. Therefore, values of semi-major axis are limited to $a=50$~AU at the start of the simulation. Generally, the evolution pattern of the $a$ in primordial binaries is similar for both simulations, that the relative frequency of small-$a$ increases with time.

 The case is different in dynamical binaries. In the \simA{} before 1~Gyr, a high fraction of dynamical binaries (Figure~\ref{fig:semi_evolution_R7_B5_solar}) have $a\lesssim 10$~AU. The fraction of dynamical binaries with $a\lesssim 10$~AU drops from 0.4 at $t=100$~Myr to to 0.1 at $t=12$~Gyr. On the contrary, in \simB, the fraction of dynamical binaries with semi-major axes $a\lesssim 10$~AU increases with time. As compared to the MS field binaries in \cite{2010ApJS..190....1R}, the semi-major axis ranges $10^{-2}-10^5$~AU, and peaks at $10-100$~AU, the \simA{} and \simB{} produced much larger number of close binaries with $a\lesssim 10$~AU, as wide binaries are less likely to survive in a star cluster environment.

%%%%%%%%%%%%%%%%%%%%%%%%%%%%%%%%%%%%%%%%%%%%%%%%%%
%%%%%%%%%%%%%%%
% eccentricity
%%%%%%%%%%%%%%%

\subsection{Eccentricity}
\label{sec:eccentricity_evolution}

\begin{figure*}
  \centering
  \includegraphics[width=0.97\textwidth]{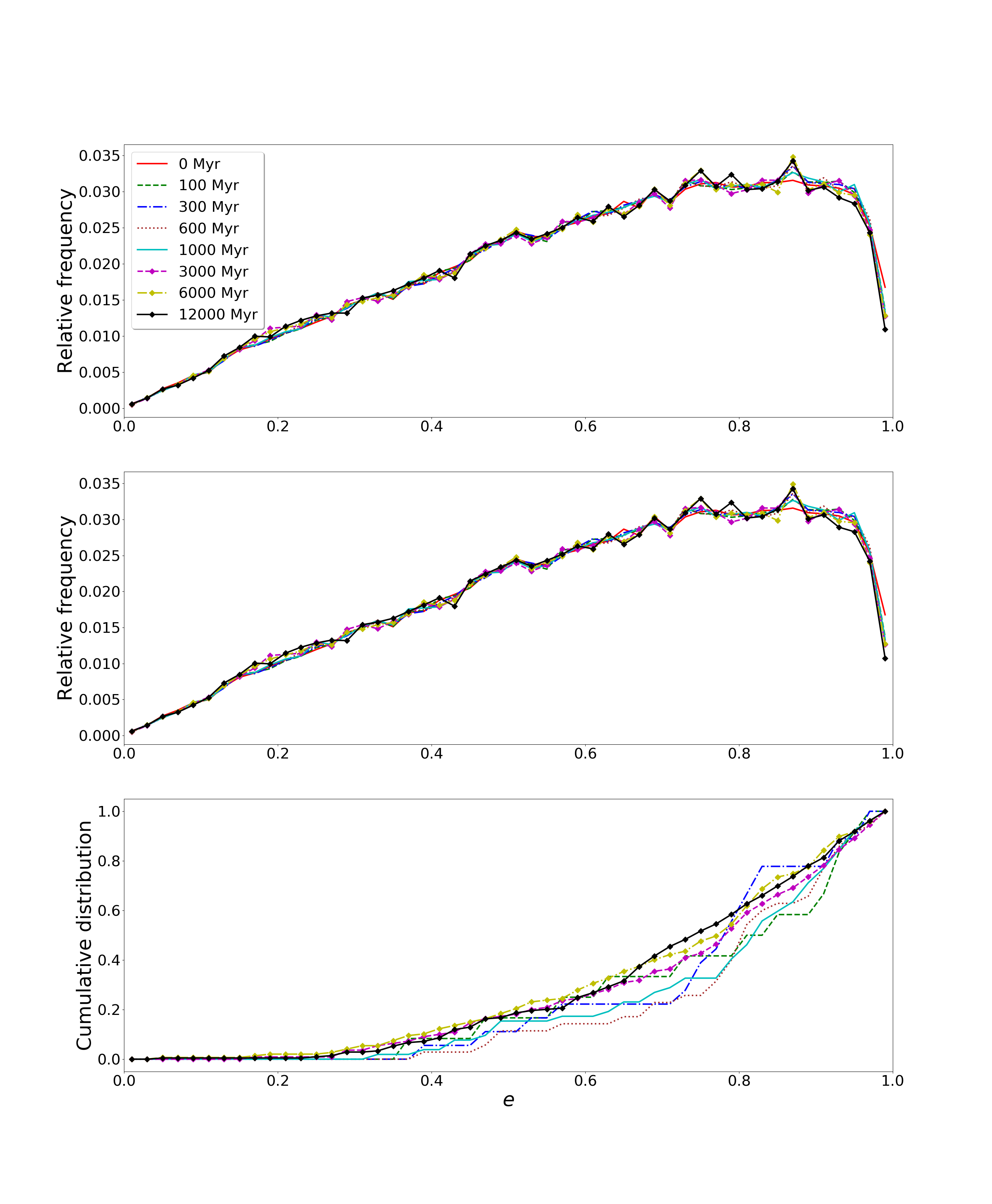}
  \caption{Secular evolution of the eccentricity $e$ of MS binaries in \simA. The colors and symbols are identical to those in Figure~\ref{fig:mass_ratio_evolution_R7_B5_solar}. }\label{fig:ecc_evolution_R7_B5_solar}
\end{figure*}

\begin{figure*}
  \centering
  \includegraphics[width=0.97\textwidth]{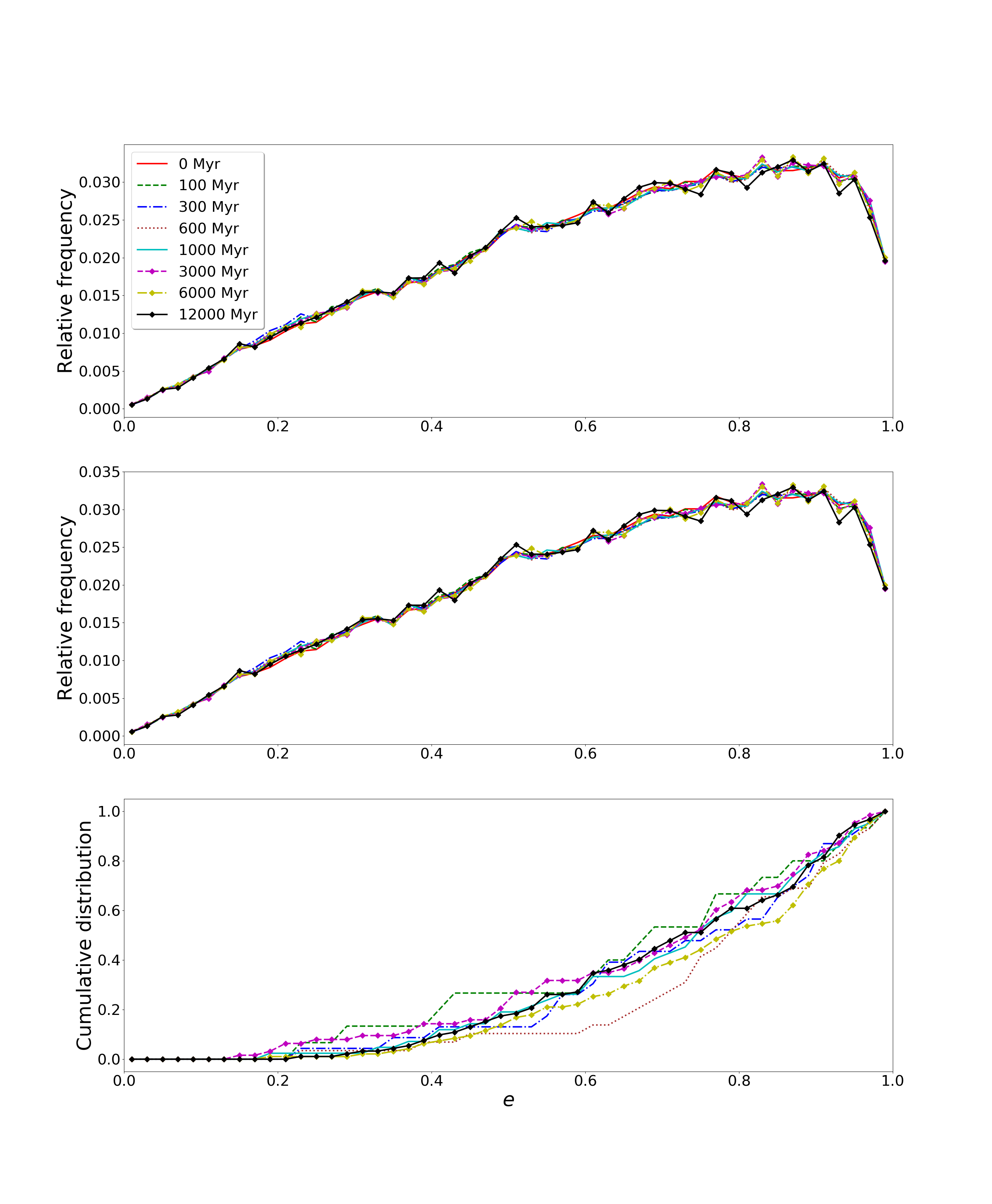}
  \caption{Secular evolution of the eccentricity $e$ of MS binaries in \simB. The color and symbols are identical to those in Figure~\ref{fig:mass_ratio_evolution_R7_B5_solar}.}\label{fig:ecc_evolution_R7_IMF2001_RG71}
\end{figure*}

Although the evolution of the semi-major axis is closely linked to that of the eccentricity $e$, the secular evolution of eccentricity is not as apparent as that of the semi-major axis and the mass ratio. 
We show the secular evolution of the eccentricity distribution, $f(e)$, in Figure~\ref{fig:ecc_evolution_R7_B5_solar} (for \simA) and Figure~\ref{fig:ecc_evolution_R7_IMF2001_RG71} (for \simB).
Both simulations \simA{} and \simB{} initially adopted a thermal eccentricity distribution, $f(e) \propto e$. To avoid immediate collisions between the two components of binary systems at the start of the simulations, binary systems with very small periastron distances are not included in the DRAGON simulations. This is the reason why there is a depression at  $e \approx 1$. Through dynamical encounters, dynamical binaries reach an energy equipartition state, which leads to a thermal eccentricity distribution for dynamical binaries, similar to the earlier discovery of \cite{1919MNRAS..79..408J}. 

Our models predict that the mean eccentricity at $t=12$~Gyr is $\langle e \rangle = 0.623$ and $\langle e \rangle = 0.627$ for models \simA{} and \simB{}, respectively. \citet{2016MNRAS.455.3009M} find a mean eccentricity of binaries within the half-mass radius (excluding the core region) of Galactic globular clusters is $e = 0.35 \pm 0.18$ \citep{2016MNRAS.455.3009M}, assuming a binary fraction of 33\%. 
The discrepancy between the two may be explained by the choice for the initial conditions of \simA{} and \simB{}, notably the initial thermal eccentricity distribution, and the initial binary fraction of 5\%. However, it should be noted that in \cite{2016MNRAS.455.3009M} only one of the binary systems is a MS-MS binary, and therefore a direct comparison has to be carried out with caution.

\subsection{Correlations between binary parameters}

In this section, we search for the correlations between the parameters of MS binaries at 12\,Gyr. We  analyse the evolution of the mass ratio distributions ($q$), semi-major axis ($a$), eccentricity ($e$), period ($p$),  primary mass,  secondary mass, and  cluster-centric distances ($r$) of binaries in the  star clusters  under consideration. 

%%%%%%%%%%%%%%%
% q VS mass correlation
%%%%%%%%%%%%%%%
\begin{figure*}
\centering
  \includegraphics[width=0.9\textwidth]{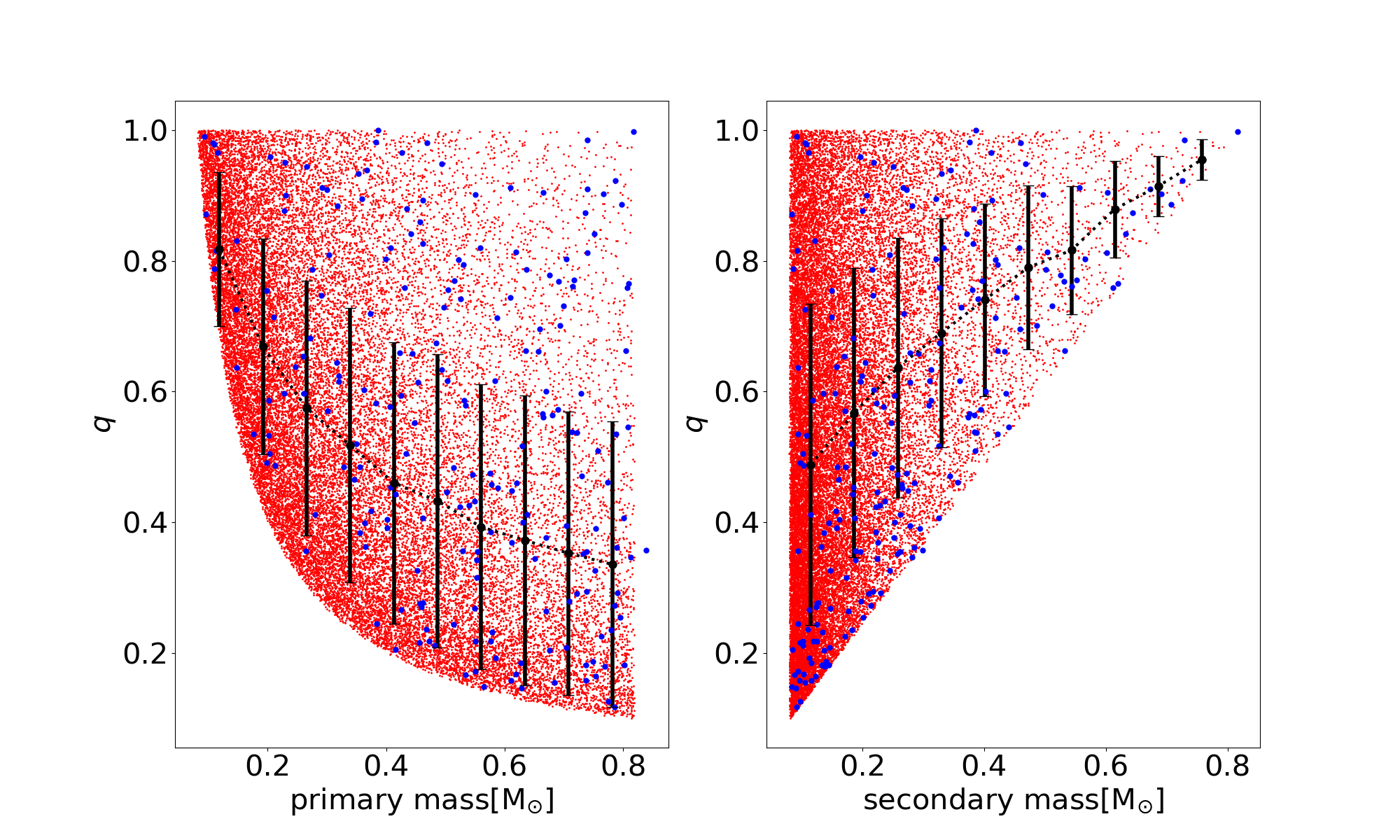}
  \includegraphics[width=0.9\textwidth]{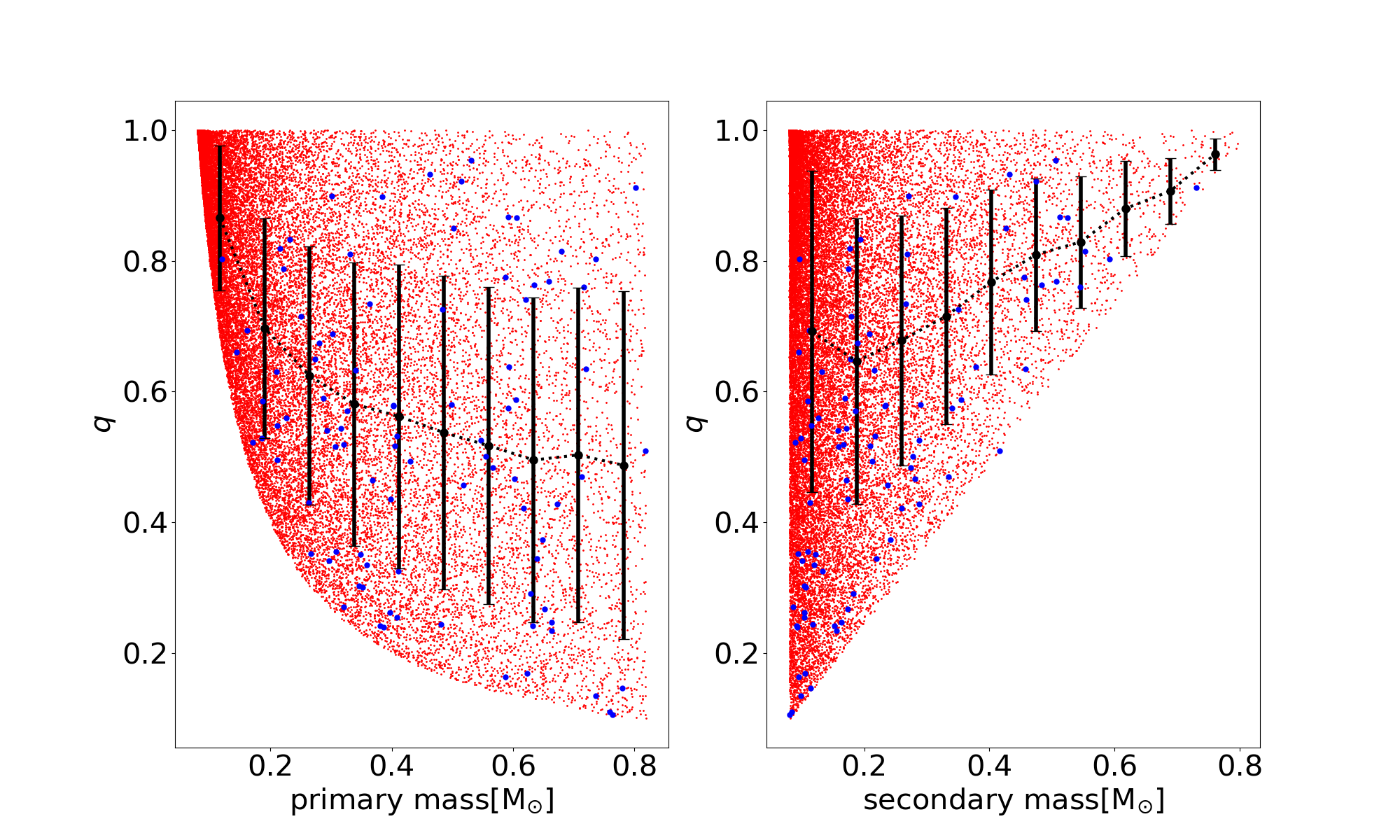} 
  \caption{The correlation between primary mass and mass ratio $q$ ({\em left}) and the correlation between secondary mass and $q$ ({\em right}), for models \simA{} ({\em top}) and \simB{} ({\em bottom}), at time $t=12$~Gyr. The red dots represent primordial binaries that remained bound throughout the entire simulation, The blue dots represent dynamical binaries that are formed at later times. The black dots mark the average mass ratio for all MS binaries for different primary masses, with the corresponding standard deviation indicated with the black error bar.
  }\label{fig:q_mass}
\end{figure*}
%%%%%%%%%%%%%%%%%%%%%%%%%%%%%%%%%%%%%

We carry out this analysis for both models \simA{} and \simB{}, and we consider both primordial binary systems and dynamical binaries.
One of the major differences between \simA{} and \simB{} is the initial binary mass ratio $q$, which generates a statistically different distributions, like shown in Figure~\ref{fig:mass_ratio_evolution_R7_B5_solar} and Figure~\ref{fig:mass_ratio_evolution_R7_IMF2001_RG71}.
We display $q$ and primary/secondary mass in Figure~\ref{fig:q_mass}. Red dots are primordial binaries, while blue dots are dynamical binaries.
There is an asymptotic boundary for the distribution of $q$ and the primary mass, which is due to the lower limit of secondary mass  0.08$\msun$,  equal to the IMF lower boundary, as $q$ is the ratio between the secondary and primary mass. 
The comparison between $q$ and the secondary mass shows a nearly straight curve with a slope of 1.25 as lower boundary, which is also the maximum possible mass of the primary star. At the end of the simulation, at 12\,Gyr, all massive star evolve over the MS. The highest mass of MS stars reaches 0.8$\msun$.

The number of dynamical binaries (blue dots) is statistically too small. The trend of total binaries (black dots) is dominated by primordial binaries (red dots). The mean mass ratio declines with increasing primary mass (black dotted curves), and as the secondary mass decrease.

Comparing black curves in \simB{} and \simA{}, the \simB{} has larger average $q$ value, because of the different initial mass ratio distribution (see Section~\ref{sec:DRAGON_simulation}).
The latter generated relatively more high-$q$ binaries, and thus larger average $q$ value.

%%%%%%%%%%%%%%%
% a e p r correlation
%%%%%%%%%%%%%%%
\begin{figure*}
\centering
  \includegraphics[width=0.8\textwidth]{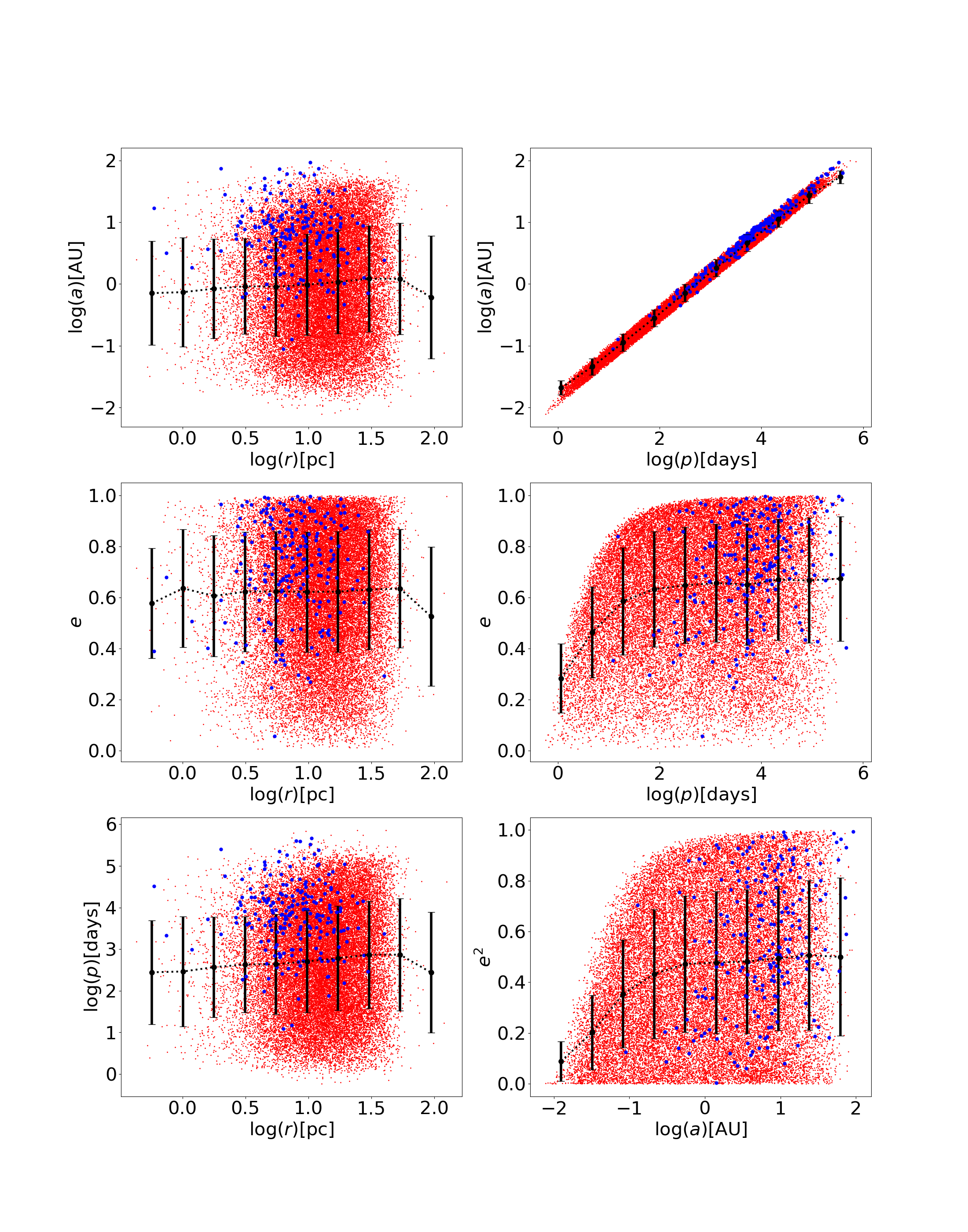}
  \caption{The correlations for parameters of main-sequence binary in model \simA{} at simulation time $t=12$~Gyr. The parameters are semi-major axis ($\log a$), eccentricity($e$ or $\log e$), orbital period ($\log p$) and cluster-centric distance ($\log r $). The red dots represent primordial binaries and the blue dots dynamical. The black dots represent the average value of $q$ for all binaries in each bin of $q$, the black error bar is the standard deviation of $q$ in each bin.
  }\label{fig:R7_B5_correlation}
\end{figure*}
%%%%%%%%%%%%%%%%%%%%%%%%%%%%%%%%%%%%%

%%%%%%%%%%%%%%%
% a e p r correlation
%%%%%%%%%%%%%%%
\begin{figure*}
\centering
  \includegraphics[width=0.8\textwidth]{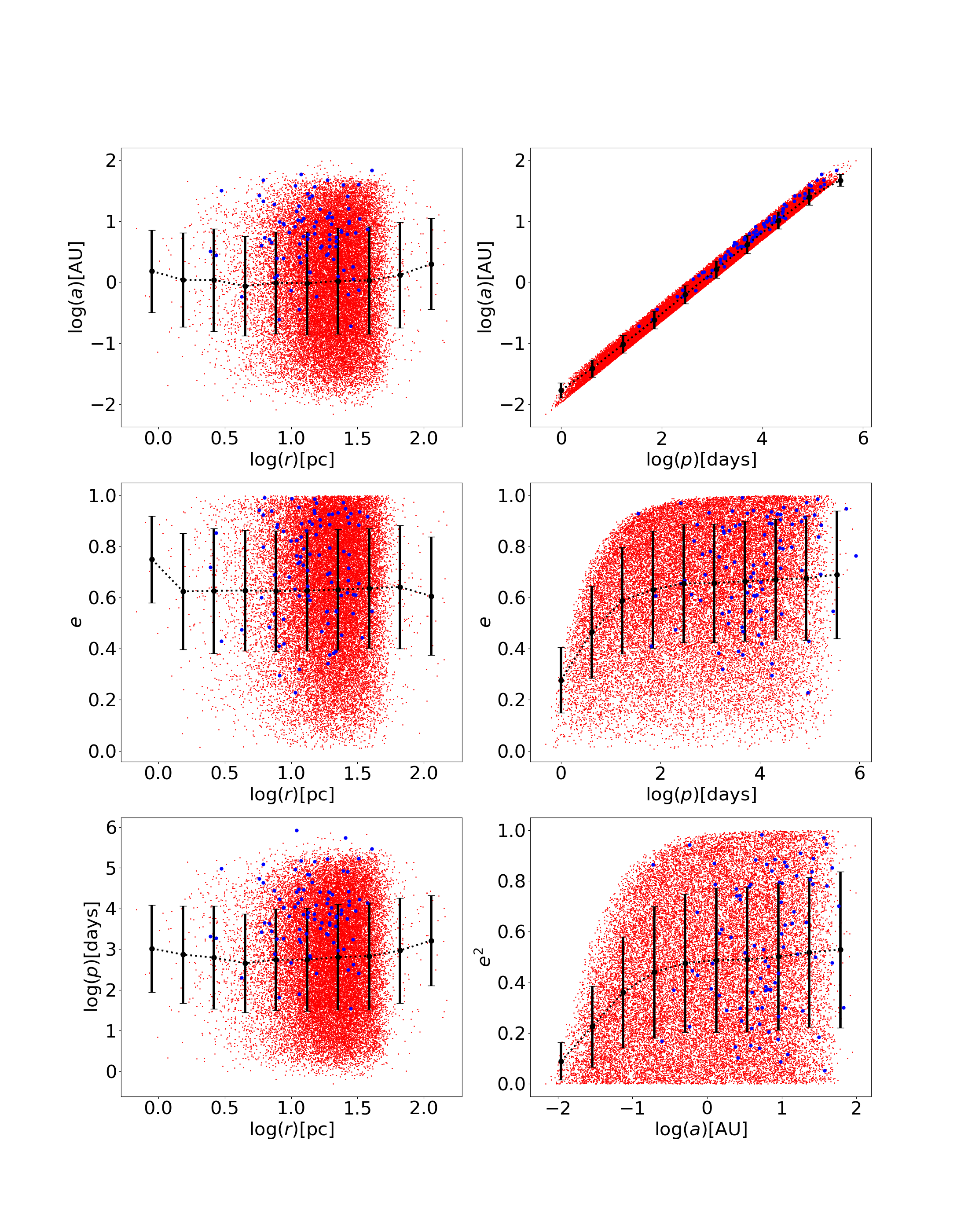} 
  \caption{The correlations for parameters of main-sequence binary in model \simB{} at simulation time $t=12$~Gyr.
  The colors and symbols are identical to those in Figure~\ref{fig:R7_B5_correlation}.
  }\label{fig:R7_IMF2001_correlation}
\end{figure*}
%%%%%%%%%%%%%%%%%%%%%%%%%%%%%%%%%%%%%

Figures~\ref{fig:R7_B5_correlation} and~\ref{fig:R7_IMF2001_correlation} show another set of correlations between parameters in \simA{} and \simB{} at 12~Gyr.
Despite the different mass ratio distribution of binaries on the initial conditions, the MS binary parameters correlations in these two simulations show very similar behaviours. We find no statistical difference of these six correlations in two simulations. The evolution of these parameters are not influenced by mass ratio. Although the considered binary are only MS stars, they form the vast majority of the binary population at all times. Therefore, the behaviour of the general binary population is well described with the MS binaries.

Generally, binaries located in the outskirts of the star clusters typically have larger semi-major axis (upper left panels in  Figure~\ref{fig:R7_B5_correlation} and ~\ref{fig:R7_IMF2001_correlation}), and longer orbital periods (bottom left panels) than their counterparts in the inner part. The linear relation between the semi-major axis and the period is simply a consequence of  Kepler's third law. The dynamical binaries (blue dots) preferentially have larger semi-major axis, longer period and more eccentric orbits than the average primordial binaries. 
The encounter probability in the inner region of the star cluster is far larger than that in the outer region, therefore the dynamical binaries form easier in the inner region. The blue dots in the left panels shows that the dynamical binaries are more, statistically, located in the inner regions than primordial binaries. The eccentricity of the system (middle left panels) does not depend on the position of the binary system in the star cluster. The values of eccentricity are related to the strength of encounters, which does not depend on the position either \citep[e.g.][]{Spurzem:2009aa, 2019MNRAS.489.2280F}.

Binary stars with small semi-major axis or short orbital periods do not have large eccentricities (middle right panels), because two very close stars are more likely to have mass transfer through the Roche-lobe overflow \citep{eggleton1983}, and circularize and/or merge shortly after. Due to initial thermal eccentricity distribution, the mean eccentricity is $e \approx 0.618$, and the distribution in $e^2$ is flat. The uniform number density in the distribution of $e^2$ versus semi-major axis indicates a tendency towards energy equipartition among MS binaries.

%%%%%%%%%%%%%%%%%%%%%%%%% 4-dynamics_binary %%%%%%%%%%%%%%%%%
\section{Dynamical signature vs. photometry of binaries in the star cluster}\label{sec:dynamics_signatures}

\subsection{Radial evolution of binaries}\label{sec:radial_binary}

\begin{figure*}
\centering
  \includegraphics[width=0.8\textwidth]{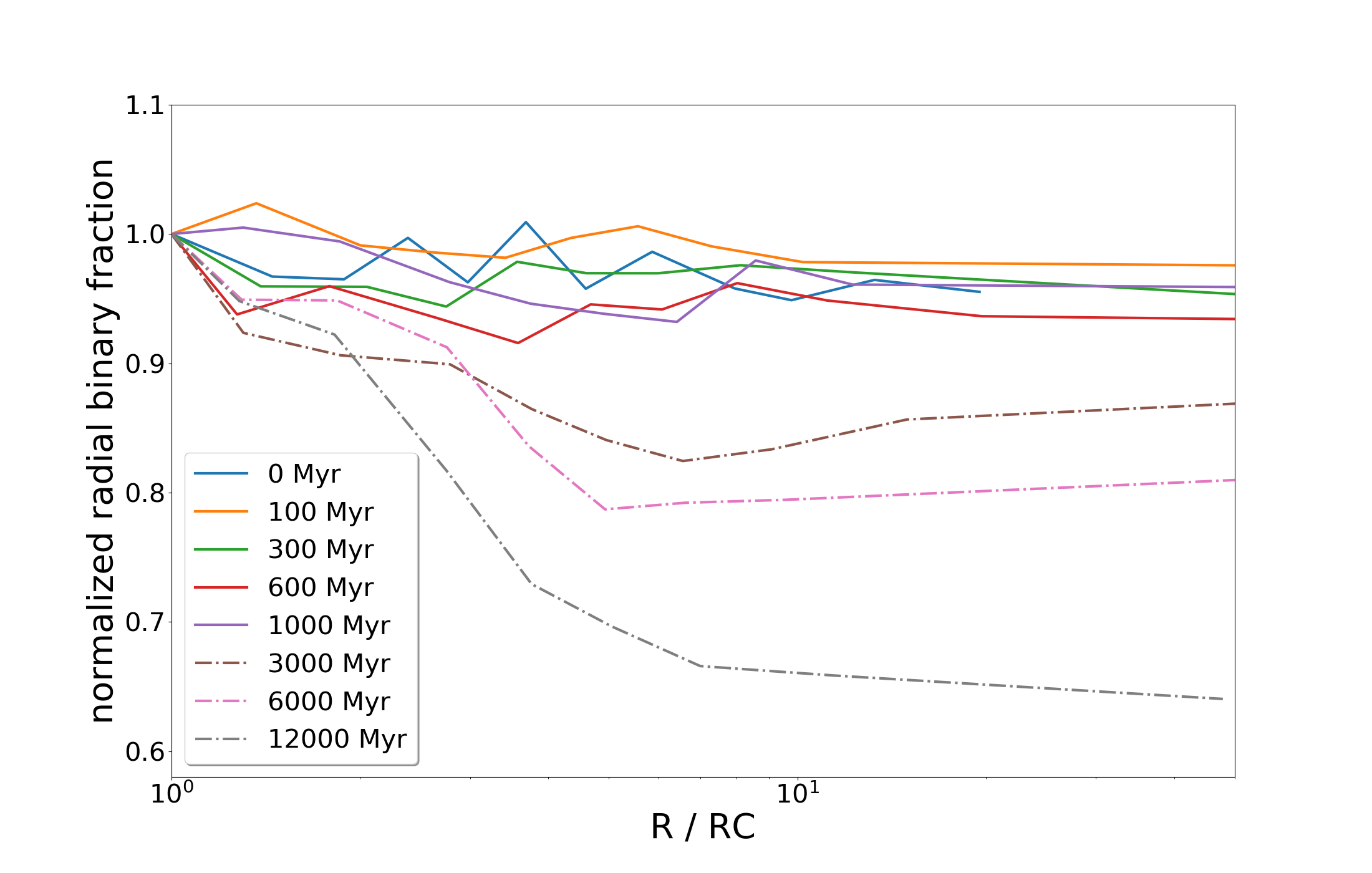}
  \includegraphics[width=0.8\textwidth]{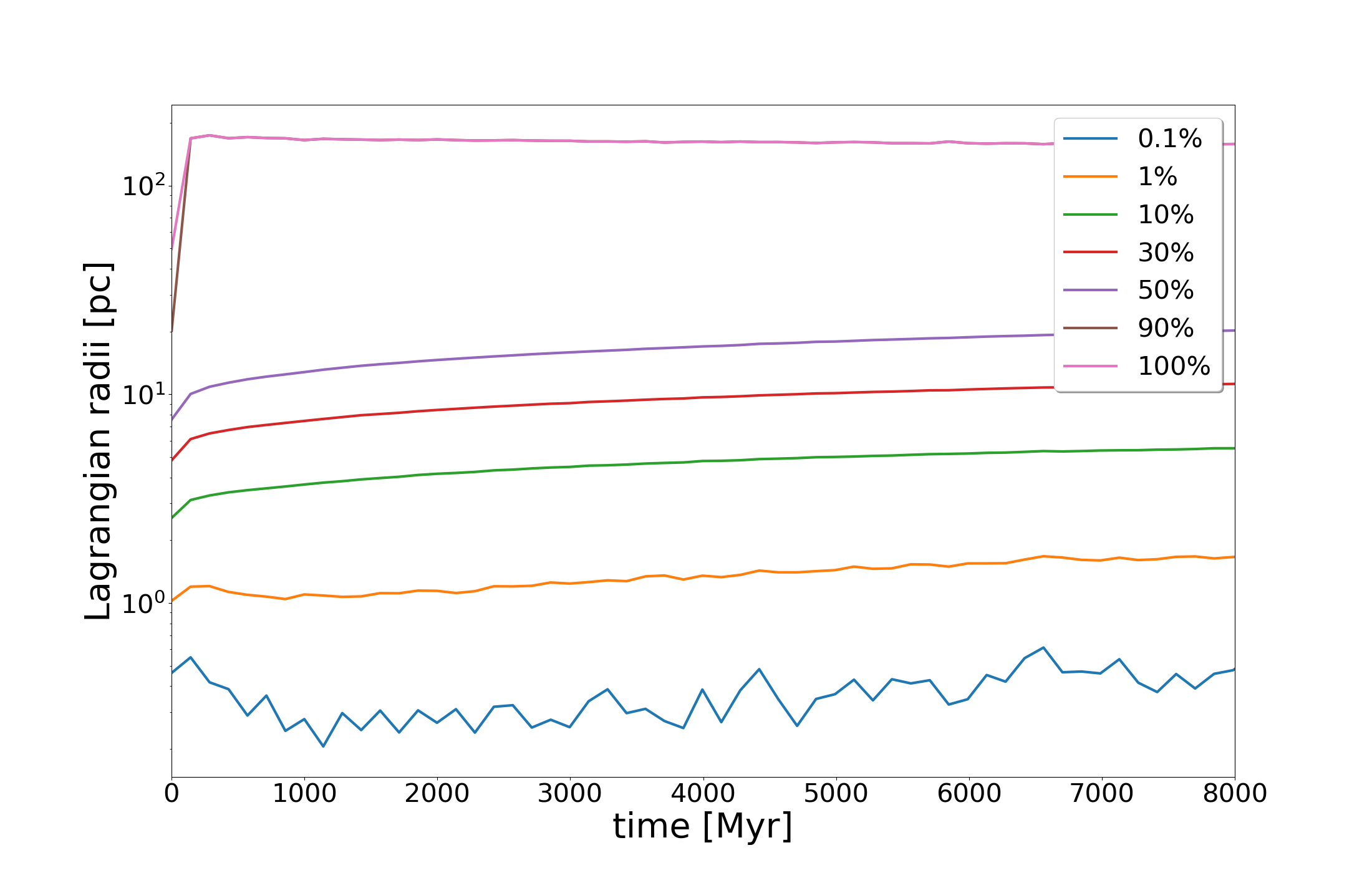}
  \caption{Radial binary fraction evolution (normalized by binary fraction in the core radius) in \simA{} ({\em top}) and Lagrangian radii evolution  in \simA{} ({\em bottom}). Lagrangian radii is frequently used in $N$-body simulation analysis, and it is defined by a certain mass fraction within a shell. For example, the 0.1\% curve shows the time evolution of the Lagrangian shell radius, which has 0.1\% of the total star cluster mass in the simulation. As the cluster expands, the Lagrangian radii for the binary systems expand similarly,  except in the special case of a core collapse, where the Lagrangian radii, in the innermost shells (such as the 0.1\% shell),  decrease.
  }\label{fig:R7_B5_solar_lagr}
\end{figure*}

\begin{figure*}
\centering
  \includegraphics[width=0.9\textwidth]{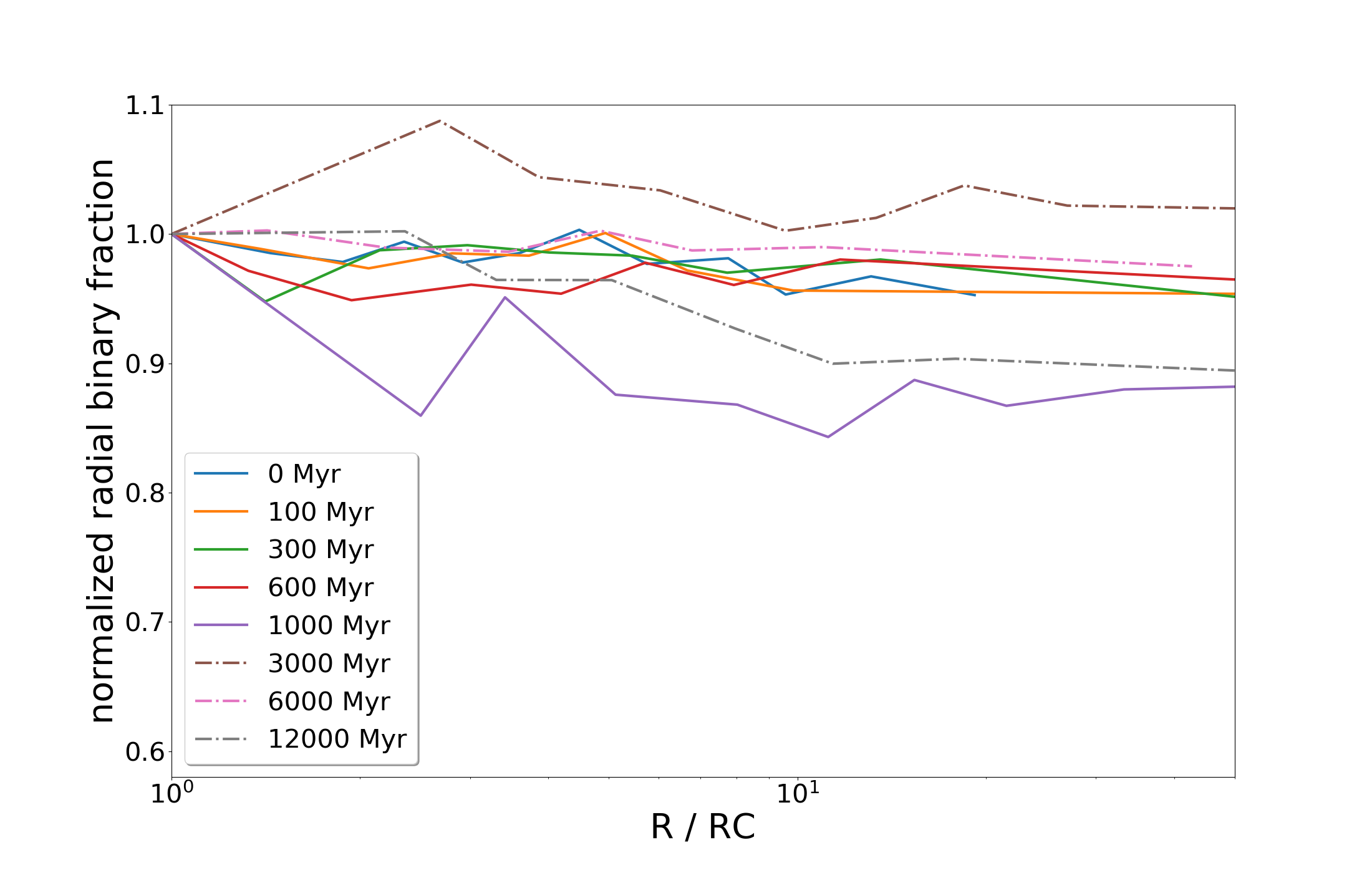}
  \includegraphics[width=0.9\textwidth]{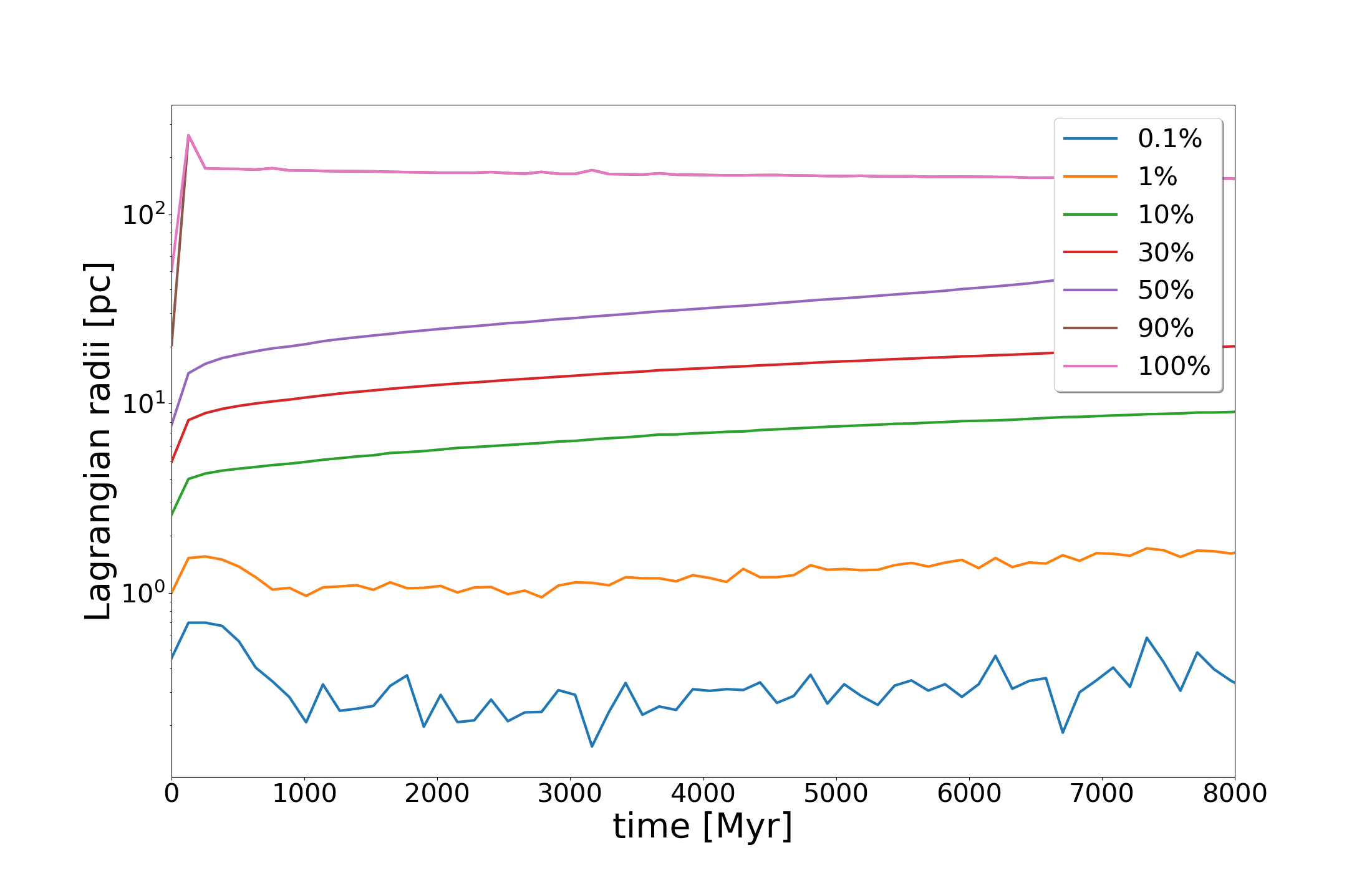}
  \caption{Radial binary fraction evolution (normalized by binary fraction in the core radius) in \simB{} ({\em top}) and Lagrangian radii evolution  in \simB{} ({\em bottom}). The colors and symbols are identical to those in Figure~\ref{fig:R7_B5_solar_lagr}.}\label{fig:R7_IMF2001_RG71_lagr}
\end{figure*}

As a consequence of the dependence of dynamical processes on the local stellar density, the binary fraction in the star cluster core tends to evolve different from that in the outskirts. 
We show the radial binary fraction, normalized to the value within the core binary fraction, in the upper panels of Figures~\ref{fig:R7_B5_solar_lagr} and~\ref{fig:R7_IMF2001_RG71_lagr}. For both models \simA{} and \simB, before 1\,Gyr, the binary fraction is larger within the core radius, and slightly drops outside the core. Only in \simA, a significant drop of binary fraction outside the core occurs after 1\,Gyr. The radial decreasing trend becomes more and more steeper with time. In model \simB, the binary fraction drops outside the core radius, and more profoundly at 1\,Gyr. Binary evolution is often closely linked to the long-term dynamical evolution of star clusters. The radial binary distribution can therefore be used as a probe of notable dynamical events that have occurred in the history of the star cluster.

In order to determine the related dynamical process at 1\,Gyr, we show the evolution of the Lagrangian radii in the bottom panel of Figures~\ref{fig:R7_B5_solar_lagr} and~\ref{fig:R7_IMF2001_RG71_lagr}. The significant decrease in the 0.1 \% and 1 \% Lagrangian radii is the evidence of core-collapse, which terminated at $t\approx 1$\,Gyr in both simulations.
The other Lagrangian radii curves show a smooth expansion, since the core-collapse affects only the innermost regions of the star cluster. The classical core-collapse phase is caused by two-body relaxation processes, which transfer energy from the inner region to the outskirts of a star cluster. 
In the DRAGON simulations, the center of the cluster is occupied by a black hole subsystem after 1\,Gyr, which is quite different from the {\bf classical paradigm.} As time passes, the radial gradient of the binary fraction increases, with the normalized radial binary fraction dropping at larger radii, as the cluster evolves approaching the time of $\sim$1\,Gyr when the core-collapse phase for both simulations is terminated. A steep radial binary fraction distribution may thus indicate a post core-collapse phase in globular clusters, and vice versa. 

\begin{figure}
\centering
  \includegraphics[width=\columnwidth]{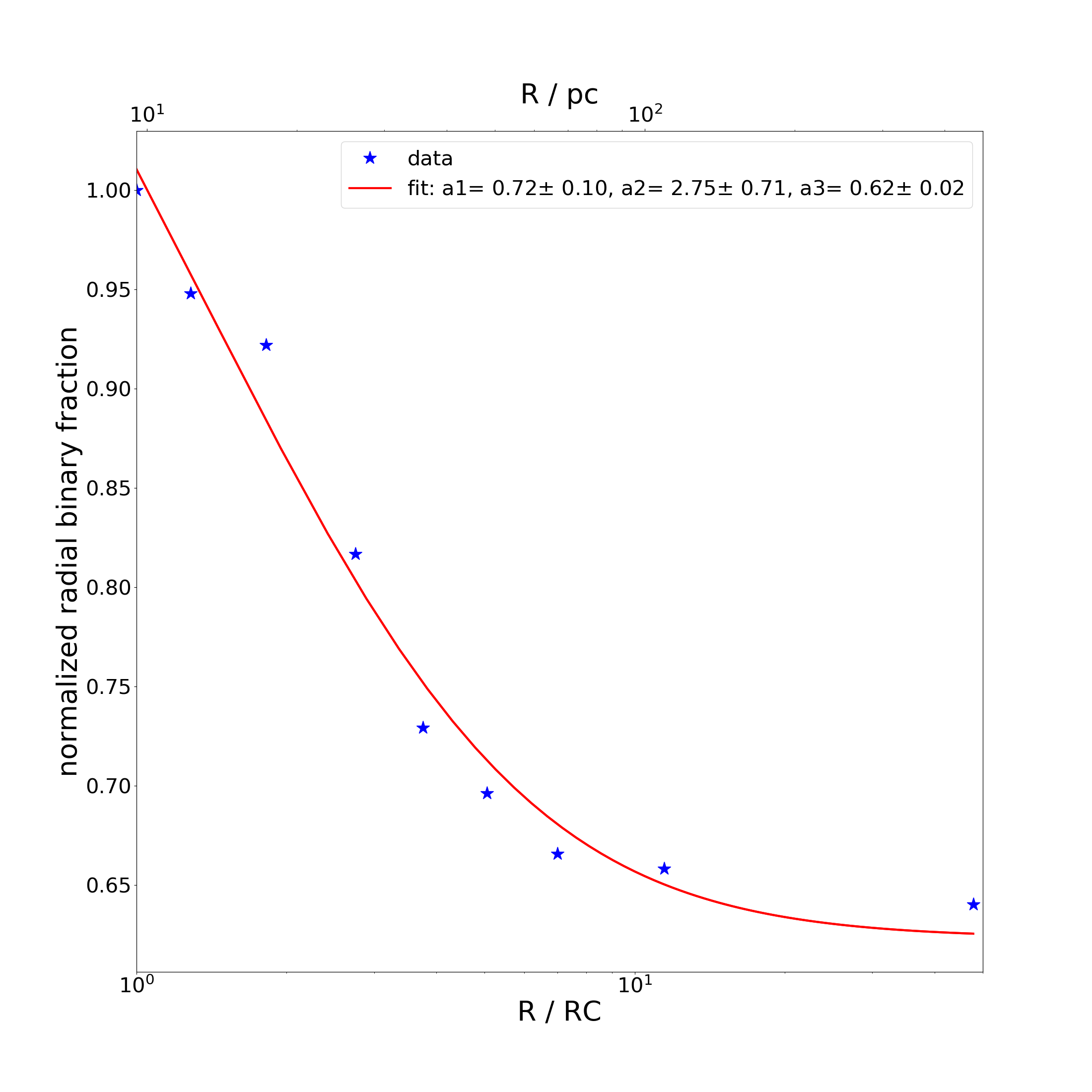}
  \caption{The normalized radial binary fraction at 12~Gyr of \simA{}  (blue asterisks). The red curve is the best fit for the blue asterisks using the analytical formula proposed by \cite{2016MNRAS.455.3009M} (see Eq.~\ref{eq:binaryfractionfit}). 
  }\label{fig:fitting}
\end{figure}

The decreasing radial binary fraction at 12\,Gyr appears to be a common trend, and is similar to the observed globular clusters \citep{2016MNRAS.455.3009M}. 
We fit the radial binary fraction distribution at 12\,Gyr of \simA{} with the function proposed by \cite{2016MNRAS.455.3009M}:
\begin{equation} \label{eq:binaryfractionfit}
    f(R_*)=\frac{a_1}{(1+R_*/a_2)^2}+a_3
    \quad ,
\end{equation}
where $R_*$ is cluster-centric distance in units of the core-radius, the same as Figure~\ref{fig:R7_B5_solar_lagr} and
Figure~\ref{fig:R7_IMF2001_RG71_lagr}.
This function generally fits the \simA{} binary sample (Figure~\ref{fig:fitting}), 
with fitted parameters of $a_1= 0.72\pm0.10$, $a_2 = 2.75\pm0.71$, and $a_3 = 0.62\pm0.02$. These values are different from the parameters of Galactic globular clusters, i.e. $a_1 = 1.05$, $a_2 = 3.5$, and $a_3 = 0.2$ \citep{2016MNRAS.455.3009M}.
The smaller value of $a_1$ in \simA{} might be due to the sample only including MS binaries. A larger value of $a_2$ indicates a steeper decrease of radial binary fraction in our sample. When all types of binaries in the simulation are included in the fit, we obtain $a_1= 0.82\pm0.10$, $a_2 = 3.04\pm0.71$, and $a_3 = 0.56\pm0.02$, which make no significant differences from the values obtained from MS binaries when fitting errors are taken into account.

%%%%%%%%%%%%%%%
% Lagrangian radii evolution
%%%%%%%%%%%%%%%

%%%%%%%%%%%%%%%%%%%%%%%%%%%%%%%%%%%%%

%%%%%%%%%%%%%%%
% Average mass evolution
%%%%%%%%%%%%%%%
\begin{figure*}
\centering
  \includegraphics[width=0.8\textwidth]{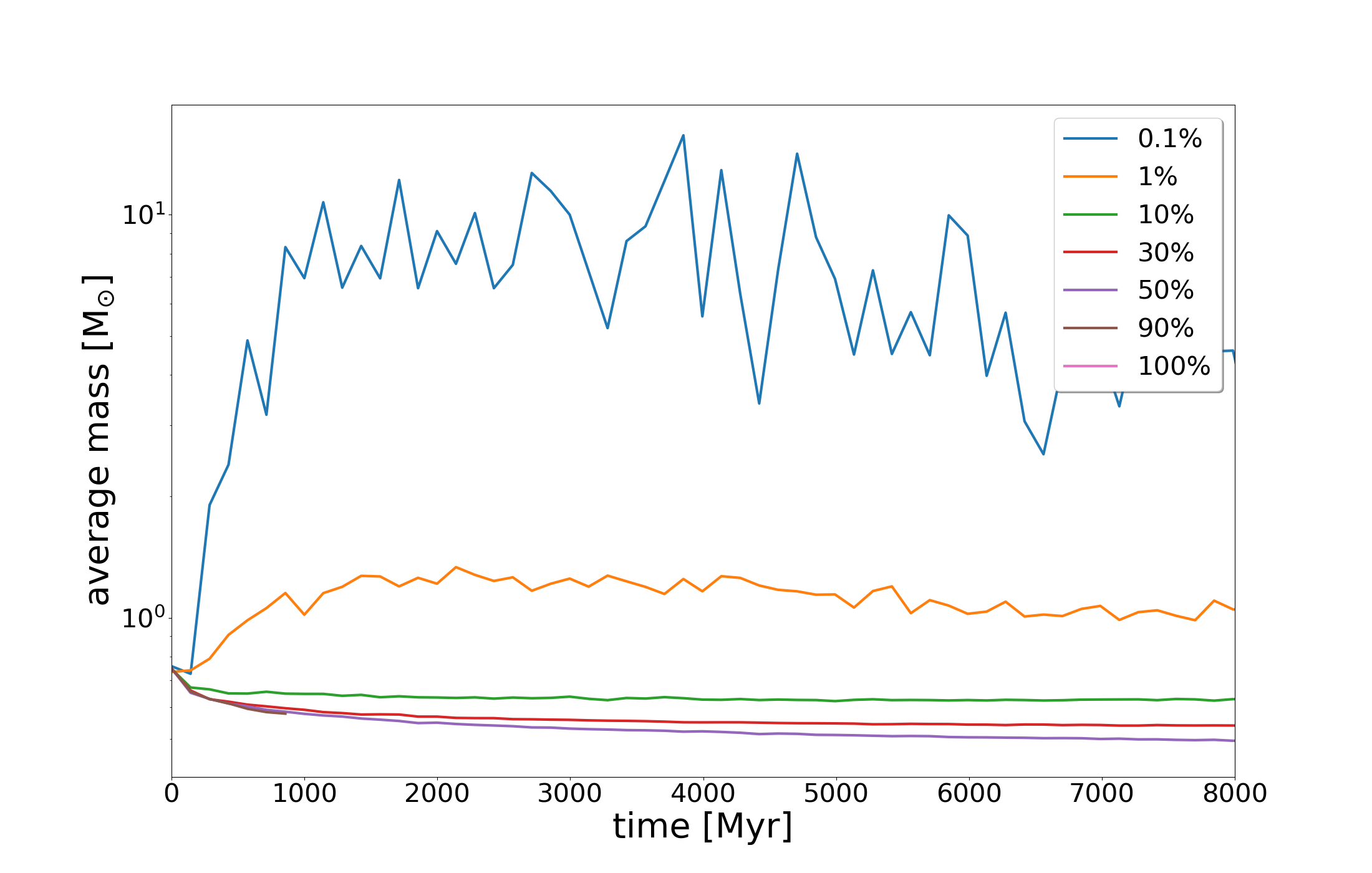}
  \includegraphics[width=0.8\textwidth]{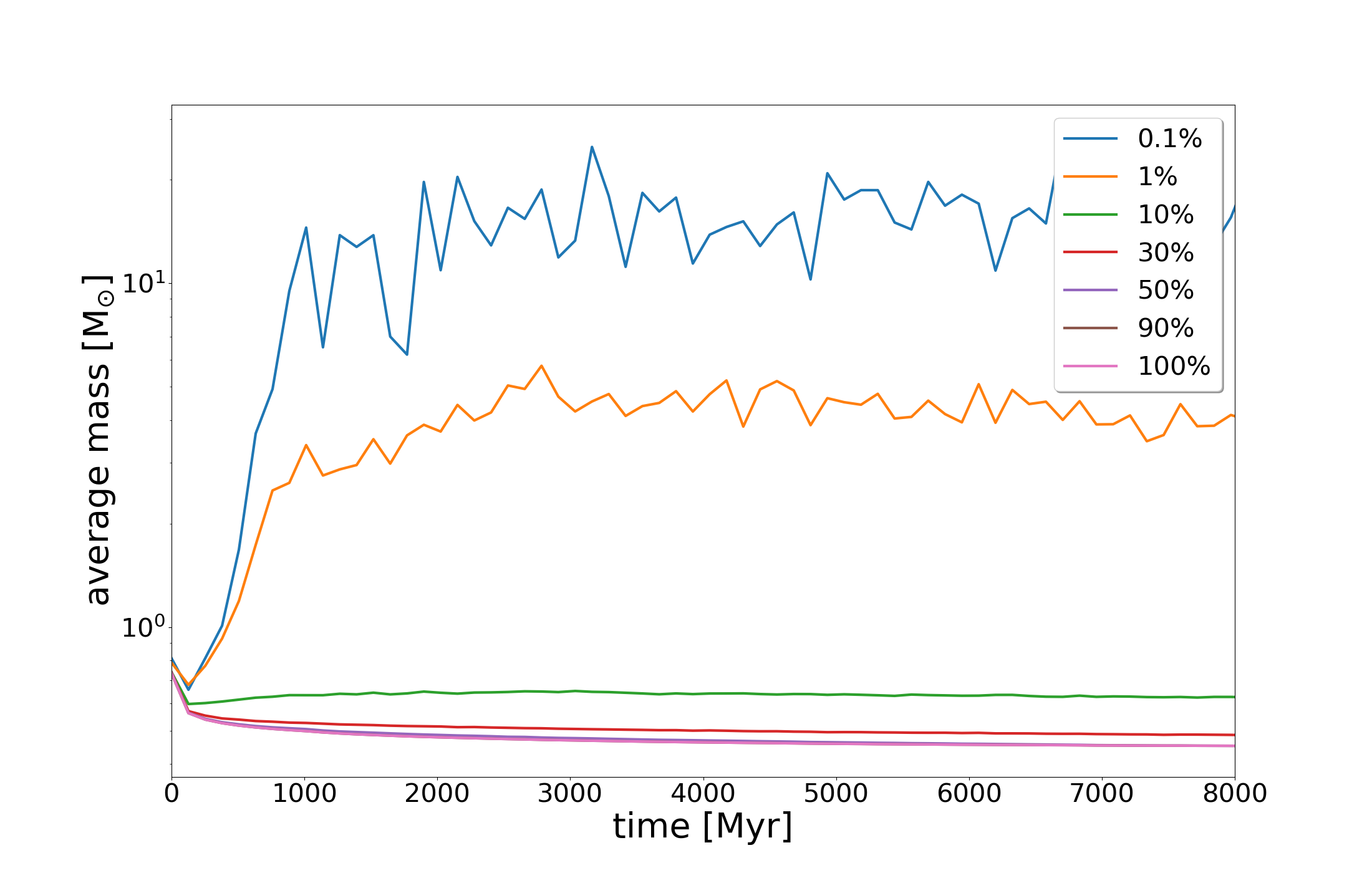} 
  \caption{Evolution of the average mass of single stars within each Lagrangian shell radius, both in \simA{} ({\em top}) and \simB{} ({\em bottom}).
  }\label{fig:avmass}
\end{figure*}

\subsection{Imprint of dynamical evolution from integrated color}\label{sec:compared_to_field}

The major dynamical process in a star cluster is two-body relaxation, which segregates massive stars to the cluster center, and low-mass stars to the outskirt  \citep{Khalisi:2007aa}.
Even a cluster as young as 1\,Myr can exhibit mass segregation \citep{2013ApJ...764...73P}, since the segregation timescale is faster for massive stars. 
The 12\,Gyr simulation time in \simA{} and \simB{}, is much longer than two-body relaxation timescale, and long enough to segregate low-mass members. We found significant mass segregation within the core in both simulations (see Figure~\ref{fig:avmass}). The mean stellar mass in the core region (within the 0.1--1\% Lagrangian radii; see Figures~\ref{fig:R7_B5_solar_lagr} and~\ref{fig:R7_IMF2001_RG71_lagr}) is more than an order of magnitude larger than in the outskirts of each cluster.
\cite{2008MNRAS.391..190G} investigated the integrated color of simulated star clusters, and found that the process of mass segregation will eventually change them. 
There was a $V-I$ color difference of roughly $0.1-0.2$\,mag between the center and outer part of the mass-segregated clusters.

\begin{figure*}
\begin{minipage}[t]{0.5\textwidth}
\centering
  \includegraphics[width=0.99\textwidth]{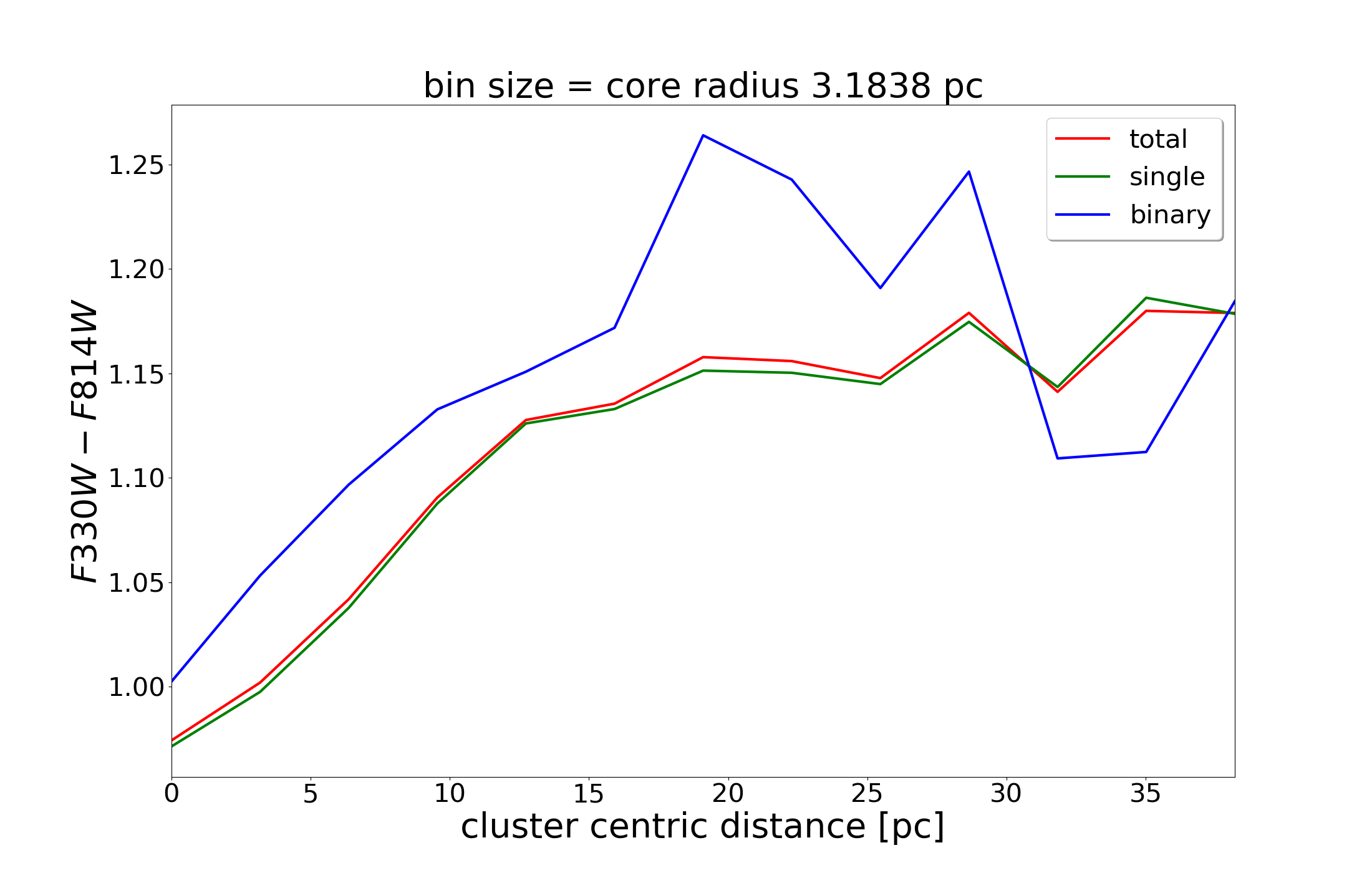}
  \includegraphics[width=0.99\textwidth]{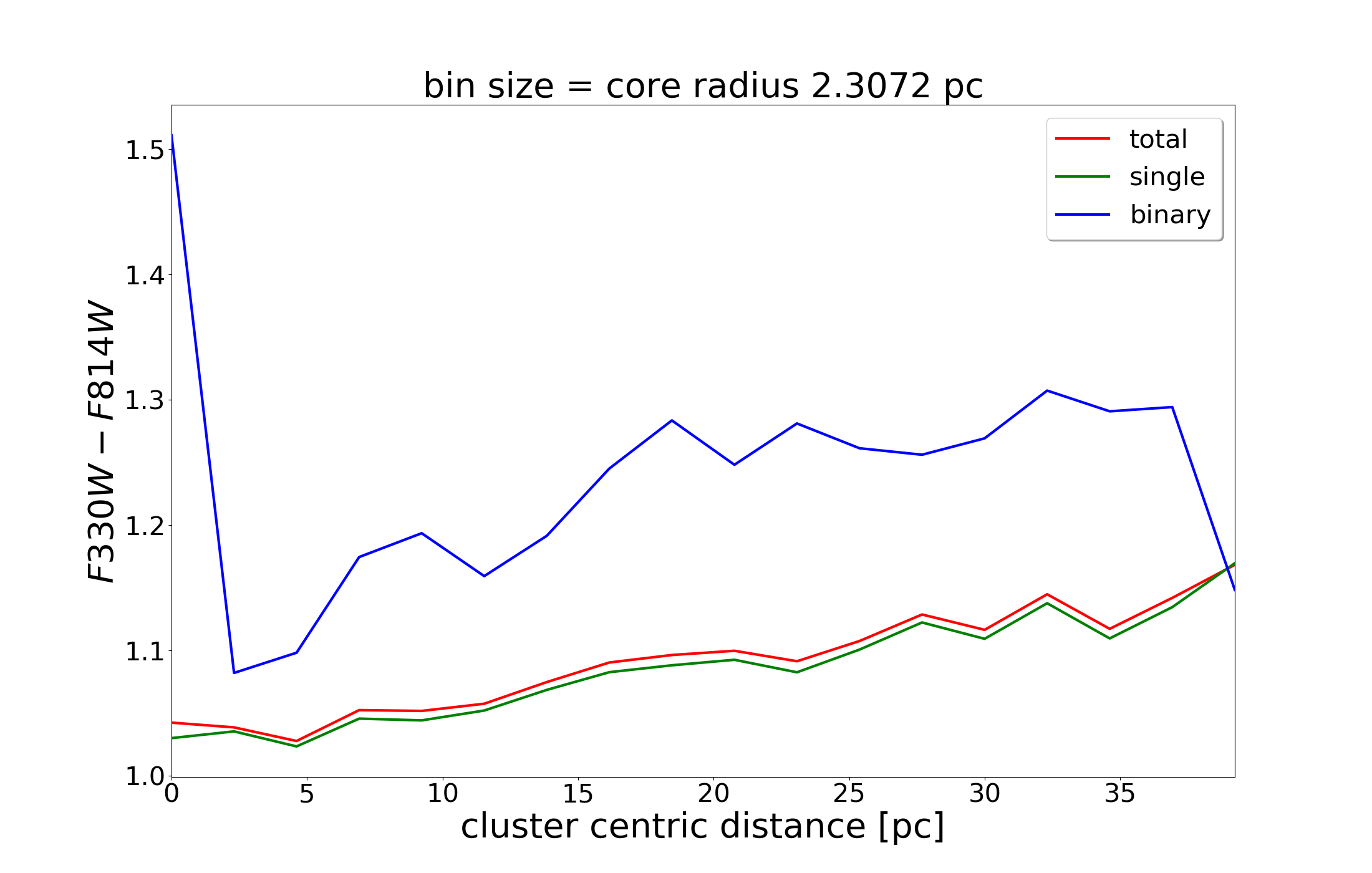}
\end{minipage}
\begin{minipage}[t]{0.5\textwidth}
\centering
  \includegraphics[width=0.99\textwidth]{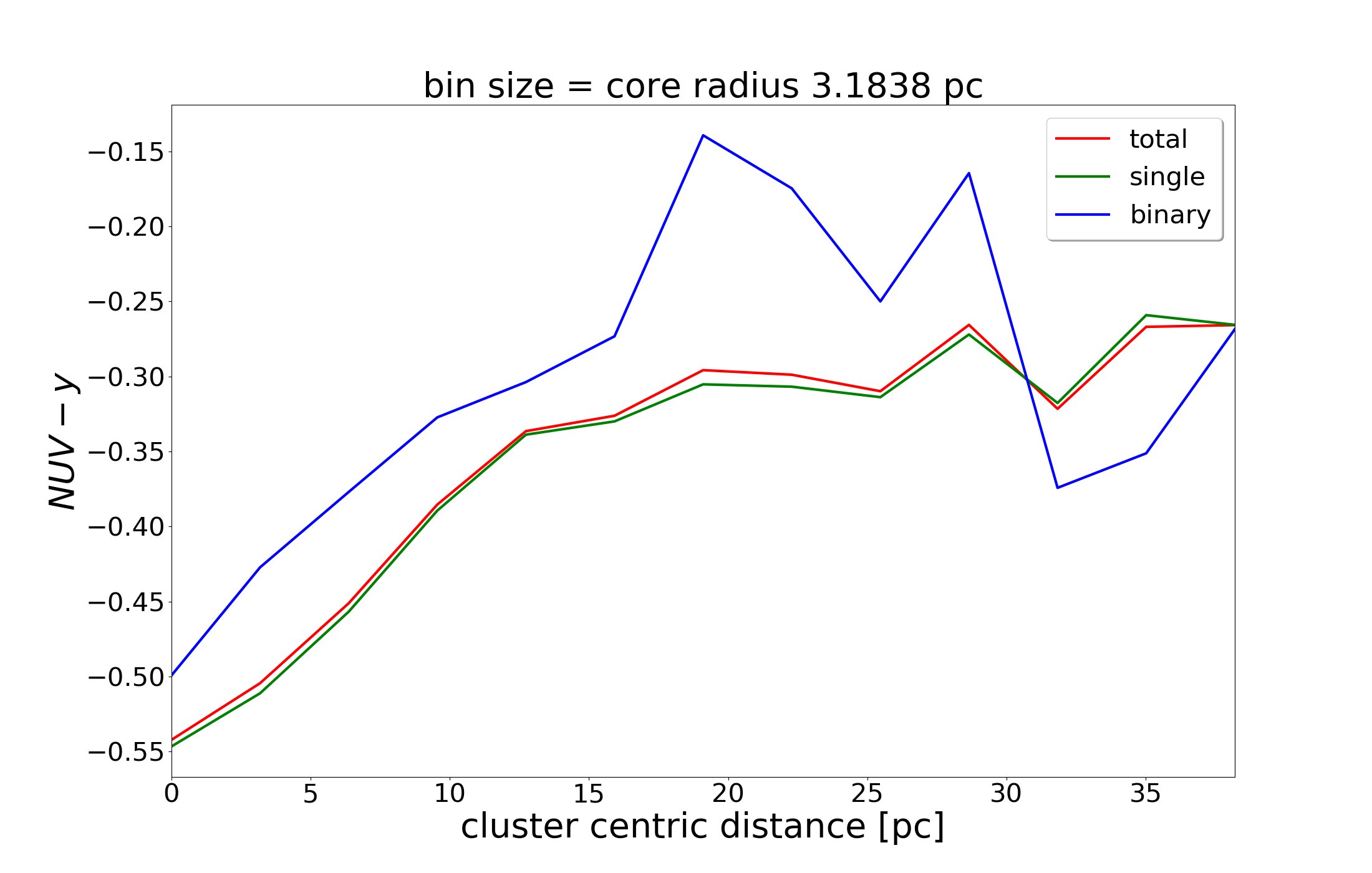} \includegraphics[width=0.99\textwidth]{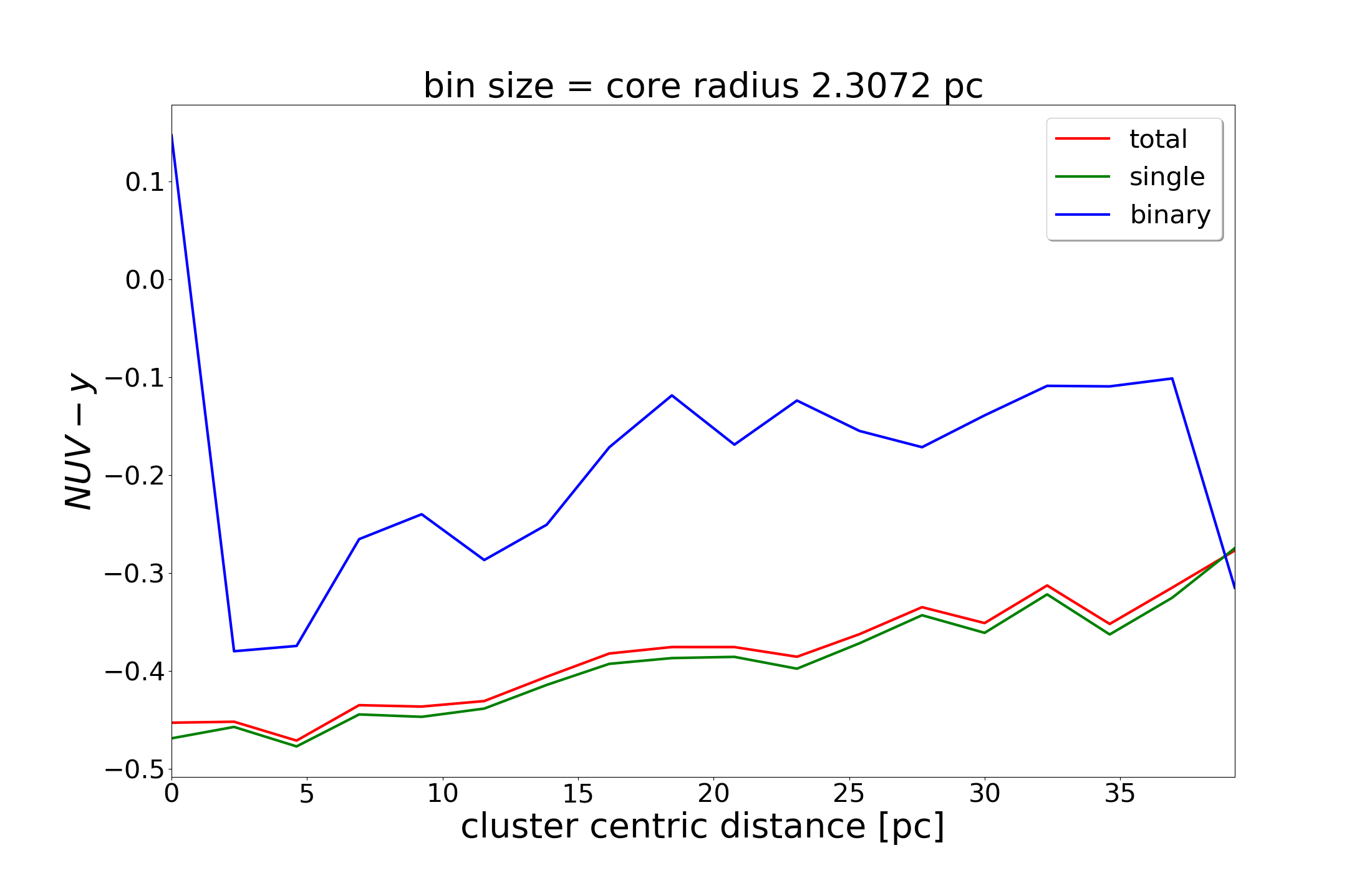}
\end{minipage}
  \caption{The color gradient of all stars (red curves), single stars (green curves) and binaries (blue curves) in \simA{} ({\em top}) and in \simB{} ({\em bottom}). The left panels use  magnitudes of HST/ACS/HRC filters, while the right CSST filters.
  }\label{fig:color_gradient}
\end{figure*}

To investigate the radial color difference in the segregated DRAGON clusters, we derived the integrated color $F330W-F814W$ ($HST$ filters) of two modeled clusters via GalevNB \citep{Pang2016} and show the color value in each annulus in Figure~\ref{fig:color_gradient}. A color gradient is found in all star and binary samples of \simA, and only in all star samples of \simB. The color is bluer in the center and redder in the outskirts, with a difference of $\sim0.1$\,mag. The result is in agreement with the findings of \cite{2008MNRAS.391..190G}. The color gradient of binary stars may reflect the steep radial profile of the binary fraction in \simA, which is not found in \simB. Thus no color gradient is found in binaries of \simB{} for the same reason. 
Therefore, mass segregation and the binary radial distribution are related. Both have an imprint in the star clusters photometry. 

The color difference of $V-I$ or $F330W-F814W=0.1$\,mag requires a photometric uncertainty of less than 0.1\,mag. This corresponds to $V (F555W)\sim23-24$\,mag in Hubble Space Telescope (HST) depending on the exposure time. HST has been the major instrument for extra-galactic star cluster observations, which are usually cannot resolved into single stars. 

The Chinese Space Station Telescope (CSST) will be an essential equipment of observing extra-galactic objects after HST.  
with a spatial resolution of $\sim 0.15''$ \citep{cao18, gong19}. 
CSST will observe down to $g=26$\,mag. 
With the ultra-deep field observation (down to 30\,mag), much better photometry is expected. In the right panels in Figure~\ref{fig:color_gradient}, we show radial color $NUV-y$ distribution for both models. The color gradient becomes more significant in $NUV-y$ color, with a color difference of $\sim0.2$\,mag, which is larger than $HST$ filters.
Therefore, CSST will be an excellent instrument used to search for the dynamical signature through radial color difference in extra-galactic star clusters.

%%%%%%%%%%%%%%%%
%% Color gradient
%%%%%%%%%%%%%%%%
%%%%%%%%%%%%%%%%%%%%%%%%%%%%%%%%%%%%%%%%%%%%%%%%

%\clearpage

%%%%%%%%%%%%%%%%%%%%%%%%% 5-Discussion_and_conclusions %%%%%%%%%%%%%%%%%%%%%%%%% 
\section{Discussion and Conclusions}\label{sec:summary}

The properties of binary systems provide important information about the formation process and dynamical history of a stellar population. In this work, we have carried out an extensive study of main sequence binaries in the DRAGON simulations, the most comprehensive direct $N$-body simulations of globular clusters available to date. We have analysed the evolution of the orbital parameters and mass properties, and compare these with observational results. In our analysis, we focus specifically on two of the DRAGON simulations: model \simA{} and model \simB{}.

The main objective of this work is to analyse the properties of the binary population in globular clusters and to make a comparison with observational data, in order to open opportunities for extracting information from simulations of star clusters. Moreover, we aim to understand the global behaviour of the population of MS  binaries in a globular cluster environments through analysing both primordial and dynamical binary systems. Our main results can be summarised as follows:

(i) For model \simA, due to the random paring binary generation method, the initial mass ratio distribution of the combined set of primordial MS binaries have relative frequency that is highest at $q \approx 0.2$. The peak value moves larger to  $q \approx 0.4$ until 12\,Gyr due to the stellar evolution. For \simB, the primordial binaries grows similarly to \simA{}, but their decrease of mass ratio is less significant, due to the different initial mass ratio distribution. \simA{} and \simB{} have similar dynamical binaries mass ratio evolution. Their dynamical binaries have at the beginning and a relatively flat mass ratio distribution at 12\,Gyr.

(ii) For both models, the semi-major axis distribution maintains a roughly uniform logarithmic distribution over time, although the binary systems with $a \ga 50$~AU  are gradually disrupted as the cluster evolves. The dynamical binary systems mostly form with semi-major axes $a \ga 1$~AU. However, these binaries are typically short-lived in a dense star cluster environment. For the \cite{Kouwenhoven2007aa} distribution, the same behaviour holds. The star cluster have, therefore, the major control over the semi-major axis of the binaries, due to the close encounters. 
We found more $a < 10$~AU binaries in our simulated star clusters (this work) than field stars \citep{2010ApJS..190....1R}. 

(iii) For both models, the eccentricity distribution of primordial binaries follows roughly a thermal distribution, but with a depression at $e \approx 1$, as a consequence of the initial conditions. This depression does not exist in dynamical binaries.

(iv) For \simA{}, the radial distribution of binary fraction is almost uniform
until 1~Gyr, After this time, the binary fraction in the outer part of the cluster tends to radially decrease. For \simB{} we do not observe a similar evolution. 

(v) MS binaries in the outskirts of clusters tend to have orbits with larger semi-major axis and longer periods than the clusters' averages. General parameter evolution of binaries is not affected by the initial mass-ratio distribution. 

(vi) Mass segregation is observed in both simulations. Moreover, the mass segregation and binary radial distribution are related. The color difference of 0.1~mag in $F330W$-$F814W$ and 0.2\,mag in $NUV-y$ from inner to outer part of the cluster, reflect the radial distribution of binaries and the mass segregation in the cluster, with similar results as \citet{2008MNRAS.391..190G}. The future Chinese Space Station Telescope (CSST) will provide an excellent opportunity to further investigate the radial color gradient in extra-galactic star clusters.

(vii) We present our data online (see Appendix~\ref{section:appendixA}). For each simulation model, we present complete sample of MS binaries at eight snapshots as $t=0$, 100, 300, 600, 1000, 3000, 6000, and 12000~Myr. This is the first public data from DRAGON project to the community. This publicly-available data resource  can be used for follow-up work and/or comparison between models. We envision that this will help to bridging computational and observational star cluster communities, and between the research groups that work on different computational approaches.

In this study we have limited our study to main-sequence binary systems, primarily because these are by far the most dominant population of binary systems in star clusters. Our focus has been primarily on two DRAGON simulations (models \simA{} and \simB{}). These two comprehensive data-sets can be used for comparison with extended simulation sets, and as benchmarks for comparison with observations in future studies. Based on the observed properties of binaries in star clusters, people can infer 
which is the proper hidden long-term evolution of binaries from our work.

%%%%%%%%%%%%%%%%%%%%%%%%% 6-Acknowledgments %%%%%%%%%%%%%%%%%
\acknowledgments
X.Y.P. is grateful to the financial support of two grants of National Natural Science Foundation of China, No. 11673032 and 11503015, and the Research
Development Fund of Xi’an Jiaotong Liverpool University (RDF-18--02--32). 
M.B.N.K. was supported by the National Natural Science Foundation of China (grant No. 11573004). F.F.D. and M.B.N.K. were supported by the Research Development Fund (grant RDF-16--01--16) of Xi'an Jiaotong-Liverpool University (XJTLU). F.F.D. acknowledges support from the XJTLU postgraduate research scholarship.
M.A.S. gratefully acknowledges Alexander von Humboldt Foundation, for financial support under the research program “Formation and evolution of black holes from stellar to galactic scales”. Part of this work benefited from support provided by the Sonderforschungsbereich SFB 881 “The Milky Way System” and the use of its GPU accelerated supercomputer.
This work has been supported by the National Natural Science Foundation of China through grant No. 11673032 (RS) and the Sino-German Cooperation Project (No. GZ1284).
 
%%%%%%%%%%%%%%%%%%%%%%%%% thebibliography %%%%%%%%%%%%%%%%%%%

\appendix
%%%%%%%%%%%%%%%%%%%%%%%%% 7-APPENDIX_A  %%%%%%%%%%%%%%%%%%%%%
\section{Binary sample}\label{section:appendixA}
We provide a complete data sample of the MS binaries in simulation model \simA{} and \simB, accessible on Github\footnote{https://github.com/qshu/table.git}  and Zenodo \footnote{https://zenodo.org/record/4042966\#.X2mb9BNLgUE}.
The file names are of the form \texttt{M\_T.table}, where \texttt{M} is the name of simulation model, and \texttt{T} is the physical time of this snapshot. We display the first ten lines of D1-R7-IMF93\_0.table shown in
Table~\ref{table:D1-R7-IMF93_0} as an example for other tables (the same format).

 \begin{sidewaystable} 
 \centering
 \caption{
         Parameters of binaries at the age of 0\,Myr in the simulation model \simA. \\ 
         The meanings of parameters are as followed. \\ 
         PDflag: primordial (1) or dynamical (0); \\ 
         $I_1$, $I_2$: name ID of these two stars in the binary, keep unchanged in different snapshots in one simulation;  \\ 
         $R_I$: distance to cluster center [pc];  \\ 
         $e$: binary eccentricity;  \\ 
         $\log(P)$: logarithmic value of binary period [days];  \\ 
         $\log(a)$: logarithmic value of binary semi-major axis [AU];  \\ 
         $M_1$, $M_2$: stellar mass of these two stars [$\msun$];  \\ 
         $\log(L_1)$, $\log(L_2)$: logarithmic value of stellar luminosity of these two stars [solar luminosity];  \\ 
         $\log(R_1)$, $\log(R_2)$: logarithmic value of stellar radius of these two stars [soalr radius];  \\ 
         $\log(T_{eff1})$, $\log(T_{eff2})$: logarithmic value of effective temperature of these two stars [K].
 }
 \label{table:D1-R7-IMF93_0}	
 \begin{tabular}{lllllllllllllll} 
 \hline
 PDflag & $I_1$ & $I_2$ & $R_I$ & $e$ & $\log(P)$ & $\log($a$)$ & $M_1$ & $M_2$ & $\log(L_1)$ & $\log(L_2)$ & $\log(R_1)$ &$\log(R_2)$ &  $\log(T_{eff1})$ &  $\log(T_{eff2})$ \\ 
 \hline
 1 & 1 & 2 & 4.767 & 0.762 & 2.985 & 0.943 & 96.385 & 0.082 & 6.113 & -3.148 & 1.010 & -0.818 & 4.785 & 3.384  \\
 1 & 3 & 4 & 12.541 & 0.803 & 4.049 & 1.603 & 68.621 & 0.109 & 5.844 & -2.694 & 0.935 & -0.859 & 4.756 & 3.518  \\
 1 & 5 & 6 & 14.910 & 0.723 & 2.276 & 0.408 & 62.280 & 0.255 & 5.763 & -1.826 & 0.912 & -0.596 & 4.747 & 3.603  \\
 1 & 7 & 8 & 8.607 & 0.888 & 3.989 & 1.544 & 59.830 & 0.329 & 5.730 & -1.634 & 0.902 & -0.519 & 4.743 & 3.612  \\
 1 & 9 & 10 & 3.005 & 0.886 & 2.143 & 0.313 & 59.346 & 0.607 & 5.723 & -0.781 & 0.900 & -0.269 & 4.742 & 3.701  \\
 1 & 11 & 12 & 4.888 & 0.452 & 3.973 & 1.527 & 56.895 & 0.717 & 5.687 & -0.443 & 0.891 & -0.193 & 4.738 & 3.747  \\
 1 & 13 & 14 & 18.930 & 0.557 & 1.401 & -0.204 & 51.291 & 0.093 & 5.598 & -2.930 & 0.867 & -0.836 & 4.728 & 3.447  \\
 1 & 15 & 16 & 16.545 & 0.897 & 3.053 & 0.889 & 47.998 & 0.552 & 5.540 & -0.963 & 0.851 & -0.315 & 4.721 & 3.679  \\
 1 & 17 & 18 & 3.097 & 0.650 & 1.991 & 0.179 & 47.674 & 0.093 & 5.533 & -2.942 & 0.850 & -0.835 & 4.720 & 3.444  \\
 1 & 19 & 20 & 7.330 & 0.640 & 3.986 & 1.508 & 47.020 & 0.565 & 5.521 & -0.918 & 0.846 & -0.304 & 4.719 & 3.684  \\
 \end{tabular}
 \end{sidewaystable}


\begin{thebibliography}{999}

\bibitem[Aarseth(1999)]{Aarseth:1999aa} Aarseth, S.~J.\ 1999, \pasp, 111, 1333

\bibitem[\protect\citeauthoryear{Allen}{2007}]{allen2007} Allen P.~R., 2007, ApJ, 668, 492

\bibitem[\protect\citeauthoryear{Allison, et al.}{2009}]{Allison:2009aa} Allison R.~J., Goodwin S.~P., Parker R.~J., de Grijs R., Portegies Zwart S.~F., Kouwenhoven M.~B.~N., 2009, ApJL, 700, L99

\bibitem[\protect\citeauthoryear{Ahmad \& Cohen}{1973}]{1973JCoPh..12..389A} Ahmad A., Cohen L., 1973, JCoPh, 12, 389 

\bibitem[Belczynski et al.(2002)]{2002ApJ...572..407B} Belczynski, K., Kalogera, V., \& Bulik, T.\ 2002, \apj, 572, 407

\bibitem[Belloni et al.(2017)]{2017MNRAS.468.2429B} Belloni, D., Zorotovic, M., Schreiber, M.~R., et al.\ 2017, \mnras, 468, 2429

\bibitem[Burgasser et al.(2003)]{2003ApJ...586..512B} Burgasser, A.~J., Kirkpatrick, J.~D., Reid, I.~N., et al.\ 2003, \apj, 586, 512

\bibitem[Cai et al.(2015)]{2015ApJS..219...31C} Cai, M.~X., Meiron, Y., Kouwenhoven, M.~B.~N., et al.\ 2015, \apjs, 219, 31

\bibitem[\protect\citeauthoryear{Catelan, Valcarce \& Sweigart}{2010}]{catelan2010} Catelan M., Valcarce A.~A.~R., Sweigart A.~V., 2010, IAUS, 266, 281, IAUS..266

\bibitem[Cao et al.(2018)]{cao18} Cao, Y., Gong, Y., Meng, X.-M., et
  al.\ 2018, \mnras, 480, 2178

\bibitem[de Zeeuw et al.(1999)]{dezeeuw1999} de Zeeuw, P.~T., Hoogerwerf, R., de Bruijne, J.~H.~J., et al.\ 1999, \aj, 117, 354

\bibitem[Eggleton(1983)]{eggleton1983} Eggleton, P.~P.\ 1983, \apj, 268, 368

\bibitem[\protect\citeauthoryear{Duch{\^e}ne \& Kraus}{2013}]{duchene2013} Duch{\^e}ne G., Kraus A., 2013, ARA\&A, 51, 269

\bibitem[Fisher et al.(2005)]{2005MNRAS.361..495F} Fisher, J., Schr{\"o}der, K.-P., \& Smith, R.~C.\ 2005, \mnras, 361, 495

\bibitem[Flammini et al.(2019)]{2019MNRAS.489.2280F} Flammini Dotti, F., Kouwenhoven, M.~B.~N., Cai, M.~X., Spurzem, R., \ 2019, \mnras, 489, 2280

\bibitem[Flammini et al.(2020)]{2020IAUS..345..293F} Flammini Dotti, F., Cai, M.~X., Kouwenhoven, M.~B.~N., Spurzem, R., Origins: From the Protosun to the First Steps of Life. Proceedings of the International Astronomical Union, 345, 293

\bibitem[\protect\citeauthoryear{Ford, Kozinsky \& Rasio}{2000}]{ford2000} Ford E.~B., Kozinsky B., Rasio F.~A., 2000, ApJ, 535, 385

\bibitem[Fujii et al.(2019)]{2019A&A...624A.110F} Fujii, M. S., Hori, Y., 2019, \aap, 624, A110

\bibitem[Gaburov \& Gieles(2008)]{2008MNRAS.391..190G} Gaburov, E., \& Gieles, M.\ 2008, \mnras, 391, 190

\bibitem[Geisler et al.(1995)]{1995AJ....109..605G} Geisler, D., Piatti, A.~E., Claria, J.~J., et al.\ 1995, \aj, 109, 605

\bibitem[Gong et al.(2019)]{gong19} Gong, Y., Liu, X.,
  Cao, Y., et al.\ 2019, \apj, 883, 203 

\bibitem[Goodman \& Hut(1993)]{1993ApJ...403..271G} Goodman, J., \& Hut, P.\ 1993, \apj, 403, 271

\bibitem[Goodwin et al.(2007)]{2007prpl.conf..133G} Goodwin, S.~P., Kroupa, P., Goodman, A., et al.\ 2007, Protostars and Planets V, 133

\bibitem[\protect\citeauthoryear{Hamers}{2020}]{hamers2020} Hamers A.~S., 2020, arXiv, arXiv:2002.08746

\bibitem[Heggie(1975)]{1975MNRAS.173..729H} Heggie, D.~C.\ 1975, \mnras, 173, 729

\bibitem[Hurley et al.(2013a)]{2013ascl.soft03015H} Hurley, J.~R., Pols, O.~R., \& Tout, C.~A.\ 2013a, SSE: Single Star Evolution, ascl:1303.015

\bibitem[Hurley et al.(2013b)]{2013ascl.soft03014H} Hurley, J.~R., Tout, C.~A., \& Pols, O.~R.\ 2013b, BSE: Binary Star Evolution, ascl:1303.014

\bibitem[Hurley \& Shara(2002)]{2002ApJ...570..184H} Hurley, J.~R., \& Shara, M.~M.\ 2002, \apj, 570, 184

\bibitem[Hurley et al.(2000)]{Hurley:2000aa} Hurley, J.~R., Pols, O.~R., \& Tout, C.~A.\ 2000, \mnras, 315, 543

\bibitem[\protect\citeauthoryear{Hurley, Tout \& Pols}{2002}]{Hurley:2002ab} Hurley J.~R., Tout C.~A., Pols O.~R., 2002, MNRAS, 329, 897

\bibitem[\protect\citeauthoryear{Hut, et al.}{1992}]{hut1992} Hut P., et al., 1992, PASP, 104, 981

\bibitem[Jeans(1919)]{1919MNRAS..79..408J} Jeans, J.~H.\ 1919, \mnras, 79, 408

\bibitem[\protect\citeauthoryear{Jiang \& Tremaine}{2010}]{jiang2010} Jiang Y.-F., Tremaine S., 2010, MNRAS, 401, 977

\bibitem[Kacharov et al.(2014)]{2014A&A...567A..69K} Kacharov, N., Bianchini, P., Koch, A., et al.\ 2014, \aap, 567, A69

\bibitem[\protect\citeauthoryear{Kalirai \& Richer}{2010}]{kalirai2010} Kalirai J.~S., Richer H.~B., 2010, RSPTA, 368, 755

\bibitem[Khalisi et al.(2007)]{Khalisi:2007aa} Khalisi, E., Amaro-Seoane, P., \& Spurzem, R.\ 2007, \mnras, 374, 703

\bibitem[King(1966)]{1966AJ.....71...64K} King, I.~R.\ 1966, \aj, 71, 64

\bibitem[\protect\citeauthoryear{Kroupa}{1995}]{kroupa1995} Kroupa P., 1995, MNRAS, 277, 1491

\bibitem[\protect\citeauthoryear{Kobulnicky, Fryer \& Kiminki}{2006}]{kobulnicky2006} Kobulnicky H.~A., Fryer C.~L., Kiminki D.~C., 2006, arXiv, astro-ph/0605069

\bibitem[Kouwenhoven et al.(2003)]{2003IAUS..221P..49K} Kouwenhoven, T., Brown, A., Gualandris, A., et al.\ 2003, IAU Symposium, P49

\bibitem[\protect\citeauthoryear{Kouwenhoven, et al.}{2005}]{kouwenhoven2005} Kouwenhoven M.~B.~N., Brown A.~G.~A., Zinnecker H., Kaper L., Portegies Zwart S.~F., 2005, A\&A, 430, 137

\bibitem[Kouwenhoven et al.(2007)]{Kouwenhoven2007aa} Kouwenhoven, M.~B.~N., Brown, A.~G.~A., Portegies Zwart, S.~F., et al.\ 2007, A\&A, 474, 77

\bibitem[\protect\citeauthoryear{Kouwenhoven et al.}{2009}]{Kouwenhoven2009aa} Kouwenhoven M.~B.~N., Brown A.~G.~A., Goodwin S.~P., Portegies Zwart S.~F., Kaper L., 2009, A\&A, 493, 979

\bibitem[\protect\citeauthoryear{Kouwenhoven, et al.}{2010}]{Kouwenhoven:2010aa} Kouwenhoven M.~B.~N., Goodwin S.~P., Parker R.~J., Davies M.~B., Malmberg D., Kroupa P., 2010, MNRAS, 404, 1835

\bibitem[Kroupa(2001)]{2001MNRAS.322..231K} Kroupa, P.\ 2001, \mnras, 322, 231

\bibitem[Kroupa(2002)]{2002MsT...........K} Kroupa, P.\ 2002, Masters Thesis

\bibitem[Kroupa et al.(1993)]{1993MNRAS.262..545K} Kroupa, P., Tout, C.~A., \& Gilmore, G.\ 1993, \mnras, 262, 545

\bibitem[\protect\citeauthoryear{Krumholz \& Thompson}{2007}]{krumholz2007} Krumholz M.~R., Thompson T.~A., 2007, ApJ, 661, 1034

\bibitem[\protect\citeauthoryear{Kustaanheimo \& Stiefel}{1965}]{Kustaanheimo:419610} Kustaanheimo P., Stiefel E., 1965, J.~Reine Angew.~Math., 218, 204 

\bibitem[\protect\citeauthoryear{Larson}{2003}]{larson2003} Larson R.~B., 2003, RPPh, 66, 1651

\bibitem[\protect\citeauthoryear{Li, et al.}{2015}]{li2015} Li Y., Kouwenhoven M.~B.~N., Stamatellos D., Goodwin S.~P., 2015, ApJ, 805, 116

\bibitem[\protect\citeauthoryear{Li, et al.}{2016}]{li2016} Li Y., Kouwenhoven M.~B.~N., Stamatellos D., Goodwin S.~P., 2016, ApJ, 831, 166

\bibitem[Makino(1991)]{1991ApJ...369..200M} Makino, J.\ 1991, \apj, 369, 200

\bibitem[\protect\citeauthoryear{Marks \& Kroupa}{2012}]{marks2012} Marks M., Kroupa P., 2012, A\&A, 543, A8

\bibitem[Mayor et al.(2004)]{2004A&A...415..391M} Mayor, M., Udry, S., Naef, D., et al.\ 2004, \aap, 415, 391

\bibitem[\protect\citeauthoryear{McKee \& Ostriker}{2007}]{mckee2007} McKee C.~F., Ostriker E.~C., 2007, ARA\&A, 45, 565

\bibitem[Meylan \& Heggie(1997)]{1997A&ARv...8....1M} Meylan, G., \& Heggie, D.~C.\ 1997, \aapr, 8, 1

\bibitem[McMillan(1986)]{1986LNP...267..156M} McMillan, S.~L.~W.\ 1986, The Use of Supercomputers in Stellar Dynamics, 156

\bibitem[\protect\citeauthoryear{Mikkola  \& Aarseth}{1993}]{Mikkola1993} Mikkola S., Aarseth S.~J., 1993, CeMDA, 57, 439

\bibitem[Milone et al.(2016)]{2016MNRAS.455.3009M} Milone, A.~P., Marino, A.~F., Bedin, L.~R., et al.\ 2016, \mnras, 455, 3009

\bibitem[Milone et al.(2010)]{2010sf2a.conf..319M} Milone, A.~P., Piotto, G., Bedin, L.~R., et al.\ 2010, SF2A-2010: Proceedings of the Annual Meeting of the French Society of Astronomy and Astrophysics, 319

\bibitem[Milone et al.(2012)]{2012A&A...540A..16M} Milone, A.~P., Piotto, G., Bedin, L.~R., et al.\ 2012, \aap, 540, A16

\bibitem[\protect\citeauthoryear{Moeckel \& Clarke}{2011}]{Moeckel:2011aa} Moeckel N., Clarke C.~J., 2011, MNRAS, 415, 1179

\bibitem[\protect\citeauthoryear{Naoz, et al.}{2013}]{naoz2013} Naoz S., Farr W.~M., Lithwick Y., Rasio F.~A., Teyssandier J., 2013, MNRAS, 431, 2155

\bibitem[Pang et al.(2016)]{Pang2016} Pang, X.-Y., Olczak, C., Guo, D.-F., et al.\ 2016, Research in Astronomy and Astrophysics, 16, 37

\bibitem[Pang et al.(2013)]{2013ApJ...764...73P} Pang, X., Grebel, E.~K., Allison, R.~J., et al.\ 2013, \apj, 764, 73

\bibitem[\protect\citeauthoryear{Price \& Bate}{2007}]{price2007} Price D.~J., Bate M.~R., 2007, MNRAS, 377, 77

\bibitem[Raghavan et al.(2010)]{2010ApJS..190....1R} Raghavan, D., McAlister, H.~A., Henry, T.~J., et al.\ 2010, \apjs, 190, 1

\bibitem[San Roman et al.(2015)]{2015A&A...579A...6S} San Roman, I., Mu{\~n}oz, C., Geisler, D., et al.\ 2015, \aap, 579, A6

\bibitem[Sana et al.(2012)]{2012Sci...337..444S} Sana, H., de Mink, S.~E., de Koter, A., et al.\ 2012, Science, 337, 444

\bibitem[\protect\citeauthoryear{Sana, James \& Gosset}{2011}]{sana2011} Sana H., James G., Gosset E., 2011, MNRAS, 416, 817

\bibitem[Shatsky \& Tokovinin(2002)]{Shatsky2002} Shatsky, N., \& Tokovinin, A.\ 2002, \aap, 382, 92

\bibitem[\protect\citeauthoryear{Shukirgaliyev, et al.}{2018}]{Shukirgaliyev2018} Shukirgaliyev B., Parmentier G., Just A., Berczik P., 2018, ApJ, 863, 171

\bibitem[Sollima et al.(2010)]{2010MNRAS.401..577S} Sollima, A., Carballo-Bello, J.~A., Beccari, G., et al.\ 2010, \mnras, 401, 577

\bibitem[Spurzem et al.(2009)]{Spurzem:2009aa} Spurzem, R., Giersz, M., Heggie, D. C., Lin, D. N. C., 2009, \apj, 697, 458

\bibitem[Tapamo(2009)]{2009LNP...791....3T} Tapamo, H.\ 2009, Jets from Young Stars V, 3

\bibitem[\protect\citeauthoryear{Tokovinin}{2014a}]{tokovinin2014a} Tokovinin A., 2014a, AJ, 147, 86

\bibitem[\protect\citeauthoryear{Tokovinin}{2000}]{tokovinin2000} Tokovinin A.~A., 2000, A\&A, 360, 997

\bibitem[\protect\citeauthoryear{Tokovinin}{2014b}]{tokovinin2014b} Tokovinin A., 2014b, AJ, 147, 87

\bibitem[Wang et al.(2015)]{Wang:2015aa} Wang, L., Spurzem, R., Aarseth, S., et al.\ 2015, \mnras, 450, 4070

\bibitem[Wang et al.(2016)]{Wang:2016aa} Wang, L., Spurzem, R., Aarseth, S., et al.\ 2016, \mnras, 458, 1450

\bibitem[\protect\citeauthoryear{Whitworth}{2001}]{whitworth2001} Whitworth A.~P., 2001, IAUS, 200, 33, IAUS..200


\end{thebibliography}
\end{document}